\documentclass[preprint,aps,eqsecnum]{revtex4-1}

\usepackage{amsmath,amssymb,amsfonts,dcolumn,color,graphicx,graphics,latexsym,placeins,epsfig}
\usepackage{subfigure,rotating,hyperref}

\newcommand{\be}{\begin{equation}}
\newcommand{\ee}{\end{equation}}
\newcommand{\ba}{\begin{eqnarray}}
\newcommand{\ea}{\end{eqnarray}}

\DeclareMathOperator{\sech}{sech}

\begin{document}

\title{\Large \bf Kinematics of trajectories in classical mechanics}

\author{Rajibul Shaikh}
\email{rajibulshaikh@cts.iitkgp.ernet.in}
\author{Sayan Kar}
\email{sayan@phy.iitkgp.ernet.in}
\author{Anirvan DasGupta}
\email{anir@mech.iitkgp.ernet.in}
\affiliation{${}^{*}$ Centre for Theoretical Studies, Indian Institute of Technology Kharagpur, 721 302, India.}
\affiliation{${}^{\dagger}$ Department of Physics, {\it and} Centre for Theoretical Studies \\ Indian Institute of Technology Kharagpur, 721 302, India.}
\affiliation{${}^{\ddagger}$ Department of Mechanical Engineering, {\it and} 
Centre for Theoretical Studies \\ Indian Institute of Technology Kharagpur, 
721 302, India.}

\begin{abstract}
\noindent In this paper, we show how the study of kinematics of a family of trajectories of a classical mechanical
system may be unified within the framework of analysis of geodesic flows in Riemannian geometry and Relativity.
After setting up the general formalism, we explore it through studies on
various one and two dimensional systems. Quantities like expansion, shear
and rotation (ESR), which are more familiar to the relativist, now re-appear
while studying such families of trajectories in configuration space, in very 
simple mechanical systems. The convergence/divergence of a family of 
trajectories
during the course of time evolution, the shear and twist of the
area enclosing the family, and the focusing/defocusing of the trajectories
within a finite time are investigated analytically for these systems. 
The understanding of the configuration space developed through such 
investigations is elaborated upon, and possible future avenues 
are pointed out.  

\end{abstract}

\maketitle

\section{\bf The approach}
\label{approach}
\noindent In classical mechanics, usually, we formulate and solve the equations of motion
of a system for certain initial conditions. In every framework, Newtonian, Lagrangian or Hamiltonian, the ultimate goal is to
develop our understanding of the system through such solutions, which
may be analytical or numerical, or a combination of both. Thus,
the focus is on the
behaviour of a {\em single} trajectory beginning from a given initial condition.
No attempt is made to study the collective behaviour of a family
of trajectories. We believe that such a study provides a new perspective
towards qualitatively understanding the configuration space of a
dynamical system.

\noindent Based on the above observation, in our work here, we frame a somewhat different question as follows:
how do we understand the behaviour of a properly defined family of
trajectories? One way of defining a family of trajectories is
to vary the initial conditions on position and velocity around specific
values.
The family of trajectories thus obtained may be treated as flow-lines in the 
configuration space of the system.
The evolution kinematics of the family may then 
be studied using the set-up discussed here. 
There are, surely, various ways of perturbing initial conditions. 
Here, we  prescribe specific initial conditions on certain kinematic
variables (that we define) which are solely associated with the family.
We analyse how the family evolves as a whole. For example, one may ask-- 
do the trajectories in the family diverge, or converge and intersect after some finite
time? If not, is there a relative shearing or a twist of the area enclosing a family of
trajectories, or do they remain parallel? Through answers to such questions, we hope 
to gain some new insights on the behaviour of a given system. 

\noindent It is natural to ask how useful is it to study
such a question. Are there any real scenarios where such an analysis is relevant?
Is there  any new formalism which can be analysed and developed?
The answer to both the queries is yes. For example, imagine two pendula,
suspended from the same point but
set to oscillate with differing initial positions and velocities. 
One may ask-- when will they strike each other? The situation 
arises, for example, in the ringing of the old-fashioned bell in temples and churches.  
Another system where the above questions 
seem relevant is projectile motion-- when and how do a bunch of projectiles 
fired with differing positions and velocities strike each other? How do the specification
of initial conditions affect the evolution of the family of projectile 
trajectories? Studies of nonlinear dynamical systems are replete
with examples where the meeting/divergence of trajectories beginning 
from two infinitesimally separated initial conditions is of interest.
It is also likely that in some situations, it may be crucial to know those
conditions which ensure that a family of trajectories never meet-- i.e., they
always tend to repel each other.
We avoid complicated systems in this work and discuss well-known, simple 
examples involving mechanical systems. The
main aim here is to illustrate the general formalism.  

\noindent The notional study on the meeting of trajectories is not new. It
has a different name and meaning in the context of Riemannian geometry and
General Relativity. It is called {\em geodesic focusing} and the
equations governing the focusing behaviour are known as the
{\em Raychaudhuri equations} \cite{akr,hawk,wald}. Its usefulness
and importance in studies related to gravitation and singularities is
well-known to all \cite{hawk,wald,kar-sengupta}. The 
Raychaudhuri equations have also been used and analysed in the 
kinematics of deformable media \cite{ADG1,ADG2} and in the kinematic 
study of geodesic congruences in various spacetime backgrounds \cite{ADG3,SG,ADG4}. However their applicability in simple 
mechanical systems has never been looked at. 
It is our hope that through the formalism and examples developed here, one will be able to 
gain useful insight into the geometrical behaviour of trajectories in 
mechanical systems.

\noindent In the next section (Section \ref{sec:one_dimension}), we develop the formalism in one 
dimension and illustrate it through some simple examples. Subsequently,
in Section \ref{sec:two_dimension}, we extend the formalism to two dimensions and
provide another set of examples-- all in two 
dimensions. In Section \ref{sec:connection}, we briefly elaborate on the links with
Riemannian geometry and Relativity. The last section (Section \ref{sec:conclusion}) 
summarises our results and proposes some questions for the future.      

\section{ One dimension}
\label{sec:one_dimension}

\subsection{Formalism}

\noindent To get started, let us consider a simple system comprising a particle of unit mass in a potential field $V(x)$ in one 
space dimension. The Lagrangian of the system is given by
\begin{equation}
L= \frac{1}{2}u^2 - V(x)
\end{equation}
where $u=\dot x$ is the velocity. The equation of motion is 
\begin{equation}
\dot u=\ddot x = -\frac{\partial V}{\partial x}
\label{eqm1}
\end{equation}
Let us now consider a family of trajectories defined by perturbing the 
initial conditions.
We introduce the variable
$\theta (t)= \frac{\partial u}{\partial x}$ which expresses the gradient of $u$
between two infinitesimally separated trajectories of the family.
Writing $\frac{{\rm d}\theta}{{\rm d}t}=u\frac{\partial\theta}{\partial x}$, we have
\begin{eqnarray}
&&\frac{{\rm d}\theta}{{\rm d}t}=
\frac{\partial}{\partial x} \left ( u\frac{\partial u}{\partial x} \right ) - \left (\frac{\partial u}{\partial x} \right )^2 \nonumber \\
&\Rightarrow& \frac{d\theta}{dt} +\theta^2 = - \frac{\partial^2 V}{\partial x^2}
\end{eqnarray}
where we have used the equation of motion (\ref{eqm1}).
Thus, given the potential function, we can obtain the equation for
$\theta$. However, in some situations, we may not know the
Lagrangian or the potential function, but know only the 
equation of motion. Examples of such systems are those 
with non-potential/non-conservative force fields (say, with damping).
In such cases, one writes $\dot u = f_{ext}$ where $f_{ext}$
includes all the forces (per unit mass) involved, including damping. We then have
\begin{equation}
\frac{d\theta}{dt} +\theta^2 = \frac{\partial f_{ext}}{\partial x}
\label{eq:1d-theta-gen}
\end{equation}
as our equation for $\theta$. We shall discuss examples of both types
in the next section.

\noindent A further important point is the notion of meeting of trajectories 
in finite time. Let us assume a general solution for $x(t)$ which depends 
on two
initial values $x_0$ and $u_0$ at $t=0$. In order to obtain a
neighbouring trajectory, one displaces $x_0$ and $u_0$ to $x_0 + \Delta x_0$
and $u_0 +\Delta u_0$, where $\Delta u_0=\frac{\partial u}{\partial x}\big\vert_{(x=x_0,t=0)} \Delta x_0=\theta_0 \Delta x_0$. Let us call the new trajectory as $x'(t)$.
At some $t=t_f$ if the two trajectories meet, then $x(t_f)=x'(t_f)$, from which
one would get the initial value of $\theta$, i.e. 
$\theta_0= \frac{\Delta u_0}{\Delta x_0}$, 
required for the trajectories to meet. Alternatively, if we assume 
$\theta_0$ we know $t_f$ without 
knowing or solving the equation for $\theta(t)$. However, the 
solution for the differential equation for $\theta$ will also yield the same
relation between the initial $\theta_0$  
and $t_f$. It may also happen that
the trajectories do not meet at all, in which case, a real finite value of $t_f$
does not exist.

\noindent In one dimension, one notices that $\theta$ represents the fractional rate of change in separation 
$(\theta=\frac{1}{\Delta x}\frac{d\Delta x}{dt})$ between two trajectories. Therefore, at any time $t$, one can write the separation $\Delta x$ as
\begin{eqnarray}
&&\int\frac{{\rm d}\Delta x}{\Delta x}=\int\theta{\rm d}t \nonumber \\
&\Rightarrow& \Delta x=\Delta x_0\;{\rm e}^{\int_0^t\theta{\rm d}t}.
\end{eqnarray}
It is clear that the (exponential) divergence/convergence of the family is intimately connected with the scalar $\theta$.
Therefore, in one dimension, one realises that $\theta$ is the well-known Lyapunov exponent \cite{strogatz}.
Further, $\theta\rightarrow -\infty\;(\infty)$ in finite time implies $\Delta x\rightarrow 0\;(\infty)$ in
finite time-- this is termed as focusing (defocussing) 
of a family/congruence of trajectories in Riemannian geometry
and Relativity. It may be noted that, at our
present level of understanding,
the identification of $\theta$ with the Lyapunov exponent is only notional.
It is not from the perspective of chaotic motion, which has to do with the
behaviour of two trajectories starting out from two slightly different initial
conditions (i.e., sensitive dependence on initial conditions).

\subsection{Examples}

\subsubsection{Simple harmonic oscillator/simple pendulum}

\noindent The equation of motion for the simple harmonic oscillator or
the simple pendulum is generically written as: 
\begin{equation}
f=\frac{d^2 x}{dt^2}=-\alpha^2 x
\label{simpleharmonic1}
\end{equation}
where, $\alpha$ is the frequency of oscillation. 
The solution of this equation is given by
\begin{equation}
x(t)=x_0 \cos(\alpha t)+\frac{u_0}{\alpha} \sin(\alpha t)
\label{eq:simpleharmonic3}
\end{equation}
and, consequently,
\begin{equation}
u(t)=-x_0\alpha \sin(\alpha t)+ u_0 \cos(\alpha t)
\label{eq:simpleharmonic4}
\end{equation}
where, $x_0$, $u_0$ are, respectively, the values of $x(t)$, $u(t)$ 
at $t=0$. 
In order to get a feel of the physical picture, let us
consider three simple pendula, hung from the same point. We take 
$x(t)$ as the angle variable for the pendula, 
with initial positions $(x_0-\Delta x_0)$, $x_0$ and $(x_0+\Delta x_0)$ and initial velocities $(u_0-\Delta u_0)$, $u_0$ 
and $(u_0+\Delta u_0)$, where $x_0$ and $u_0$ are the 
values of the initial position and initial velocity of the central 
pendulum.
The three pendula will strike
each other when the separation between them becomes zero, i.e. when
\begin{equation}  
\Delta x_0\cos(\alpha t_f)+\frac{\Delta u_0}{\alpha} \sin(\alpha t_f)=0
\end{equation}
This condition may be recast as 
\begin{equation}
\alpha+\theta_0 \tan(\alpha t_f)=0
\end{equation}
where we have defined $\theta_0=\frac{\Delta u_0}{\Delta x_0}$.
Note that $t_f$ depends on $\alpha$ and, more importantly, on $\theta_0$.
As stated earlier, this result must also emerge from the solution of the
differential equation for $\theta(t)$, which, in this example is,
\begin{equation}
\frac{d\theta}{dt}+\theta^2+\alpha^2=0
\label{eq:simpleharmonic2}
\end{equation}
The solution of the above equation is,
\begin{equation}
\theta(t)=\alpha\frac{\theta_0-\alpha \tan(\alpha t)}{\alpha+\theta_0 \tan(\alpha t)}
\label{eq:simpleharmonic5}
\end{equation}
It is easy to note from (\ref{eq:simpleharmonic5}) that 
$\theta\rightarrow -\infty$ as $t\rightarrow t_f$ which obviously indicates
that the family of trajectories intersect in finite time.

\begin{figure}[ht]
\centering
\includegraphics[scale=0.9]{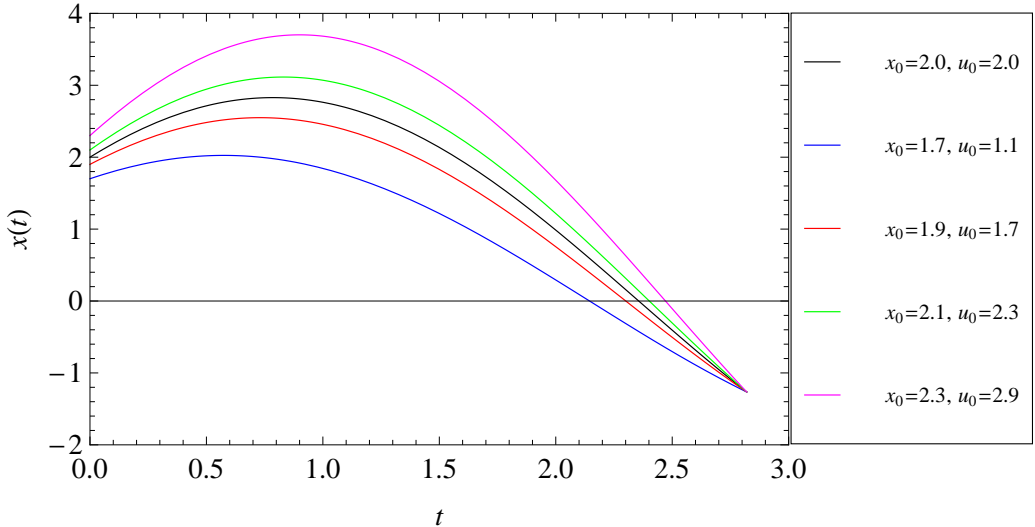}
\caption{Plots of positions of pendula having different $x_{0}$ and $u_{0}$. The separations $\Delta x_0$ and $\Delta u_0$ between the trajectories 
symmetric around the central trajectory (black curve) are chosen in such a way that they have same $\theta_0=\frac{\Delta u_0}{\Delta x_0}$. Here $\theta_0=3.0$, $\alpha=1.0$.}
\label{fig:simpleharmonic1}
\end{figure}

\begin{figure}[ht]
\centering
\subfigure[]{\includegraphics[scale=0.5]{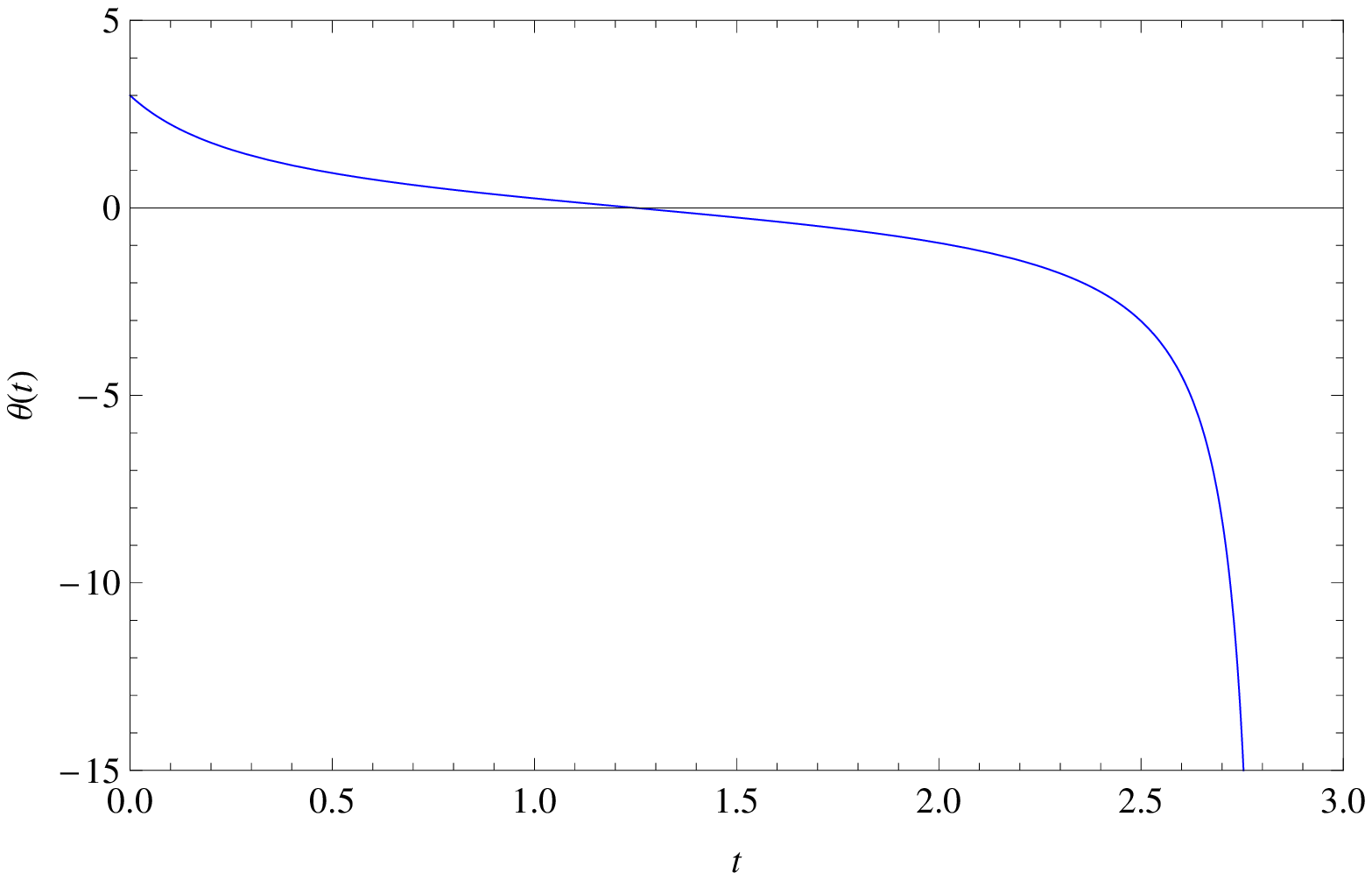}\label{fig:simpleharmonic2a}}
\subfigure[]{\includegraphics[scale=0.5]{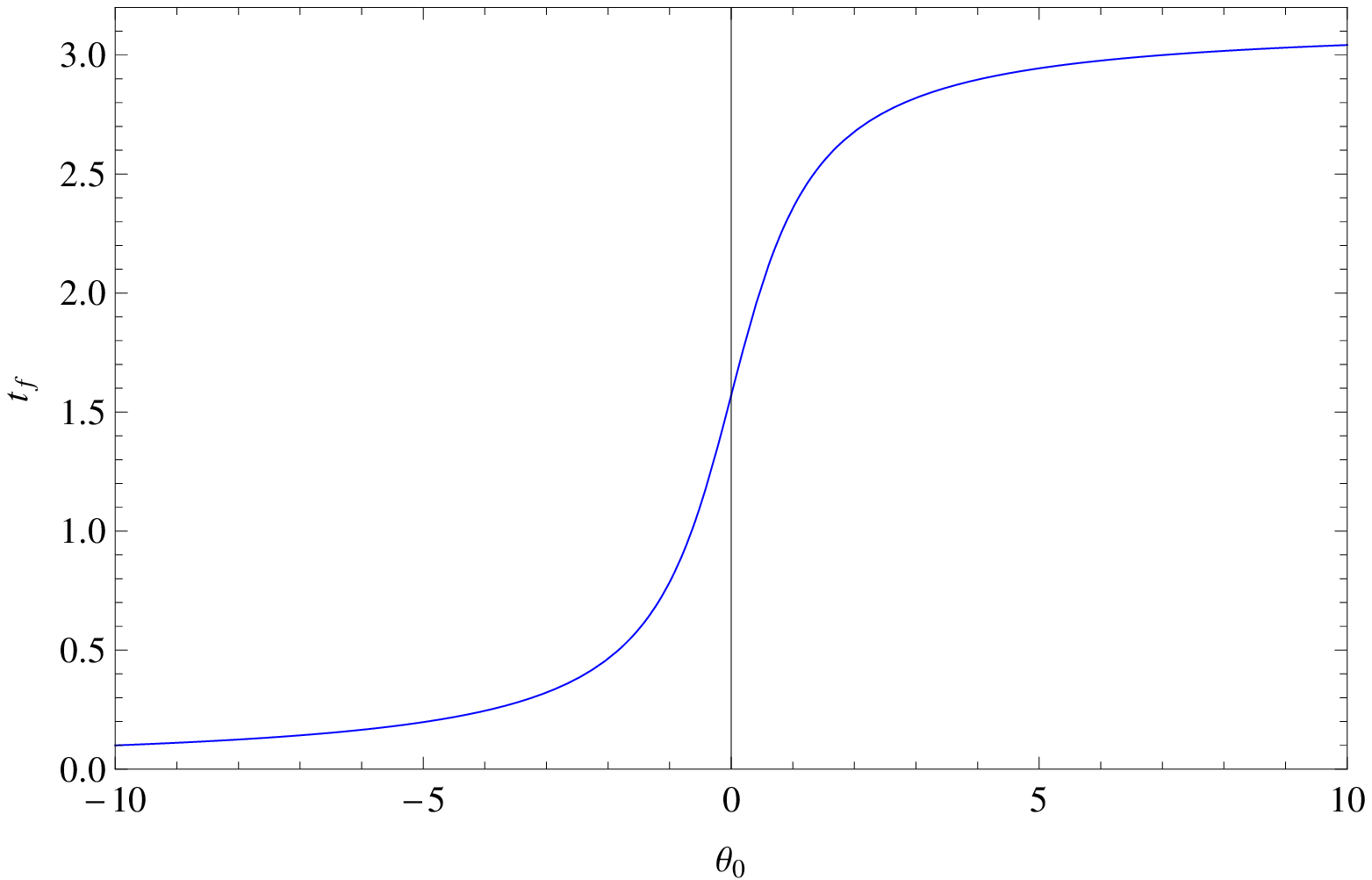}\label{fig:simpleharmonic2b}}
\caption{Plots of $(a)$ $\theta(t)$ for $\theta_0=3.0$ and $(b)$ focusing time $t_f$ for different initial $\theta_0$.}
\label{fig:simpleharmonic2}
\end{figure}

\noindent Fig.~\ref{fig:simpleharmonic1} shows the position of five pendula 
with the black one as the central pendulum. The corresponding plot for $\theta(t)$ is shown in Fig.~\ref{fig:simpleharmonic2a}. Fig.~\ref{fig:simpleharmonic2b} shows the focusing time $t_f$ as a function of initial expansion $\theta_0$.

\subsubsection{Particle falling under gravity with drag linear in speed}
\noindent According to the well-known Stokes' law, the drag force per unit mass on 
a small spherical body moving through a viscous medium is $\beta u$, where $\beta$ is a constant which depends on the radius of the particle and on the viscous coefficient of the medium. $u$ is the particle velocity. Let, the particle 
move through the medium under the action of a constant force $mg$. The equation of motion and the equation for $\theta(t)$ become
\begin{equation}
f=\frac{du}{dt}=\frac{d^2x}{dt^2}=\alpha-\beta u
\label{eq:stokes1}
\end{equation}
\begin{equation}
\frac{d\theta}{dt}+\theta^2+\beta\theta=0
\label{eq:stokes2}
\end{equation}
where $\alpha=g$. The above set of equations can be solved to obtain
\begin{equation}
x(t)=x_0+\frac{\alpha}{\beta}t+\frac{1}{\beta}\left(u_0-\frac{\alpha}{\beta}\right)(1-e^{-\beta t})
\label{eq:stokes3}
\end{equation}
\begin{equation}
\theta(t)=\frac{\beta e^{-\beta t}}{\frac{\beta}{\theta_0}+(1-e^{-\beta t})}
\label{eq:stokes4}
\end{equation}
where, $x_0$, $u_0$ and $\theta_0$ are respectively position, velocity and 
expansion at $t=0$.

\begin{figure}[ht]
\centering
\includegraphics[scale=1.0]{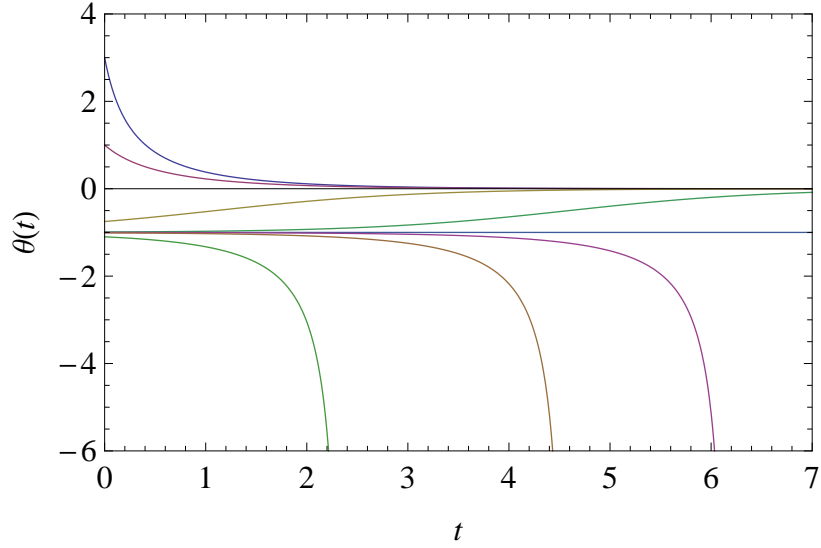}
\caption{Plots of $\theta(t)$ for different $\theta_0$. Here, $\alpha=1$, $\beta=1$.}
\label{fig:stokes1}
\end{figure}

\begin{figure}[ht]
\centering
\includegraphics[scale=0.90]{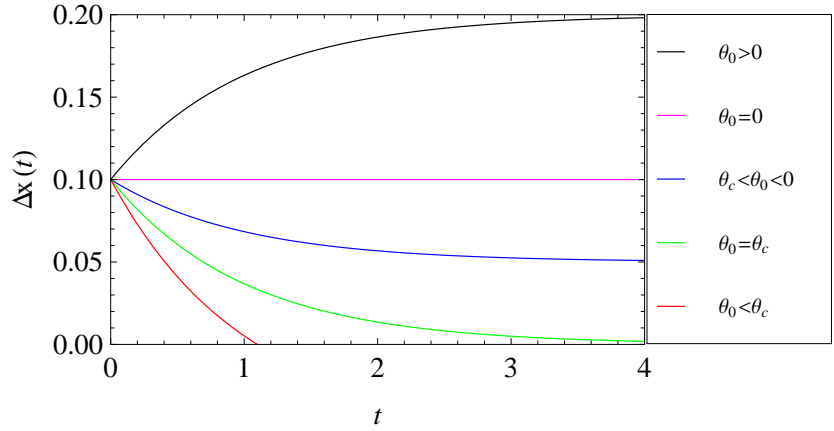}
\caption{Plots of the separation $\Delta x(t)$ between two trajectories for 
different $\theta_0$.}
\label{fig:stokes2}
\end{figure}

\noindent Fig.~\ref{fig:stokes1} shows the plot of the expansion for different 
initial values. From the differential equation for $\theta(t)$ and also from 
the Fig.~\ref{fig:stokes1}, it is clear that $\theta_0=0$ and $\theta_0=-\beta$ are the fixed points of 
the system. Fig.~\ref{fig:stokes2} shows the separation between two trajectories 
as a function of time, for different initial expansion $\theta_0$. 
For $\theta_0<\theta_c=-\beta$, the meeting of trajectories takes place, i.e., 
separation becomes zero in finite time. For $\theta_0=\theta_c=-\beta$, 
the separation goes to zero as $t\rightarrow\infty$. For $\theta_c<\theta_0<0$, the separation decreases and is finite as $t\rightarrow\infty$. If $\theta_0=0$, the separation remains the same for all time. For $\theta_0>0$, the separation 
increases and is finite as $t\rightarrow\infty$. The interesting feature
here (unlike the previous example) is the existence of a critical $\theta_0=\theta_c$, below and above
which the behaviour of the family of trajectories is different. We also note
here that there can be situations where the trajectories never meet. 

\subsubsection{Particle falling under gravity with drag quadratic in speed}
\begin{figure}[ht]
\centering
\includegraphics[scale=0.97]{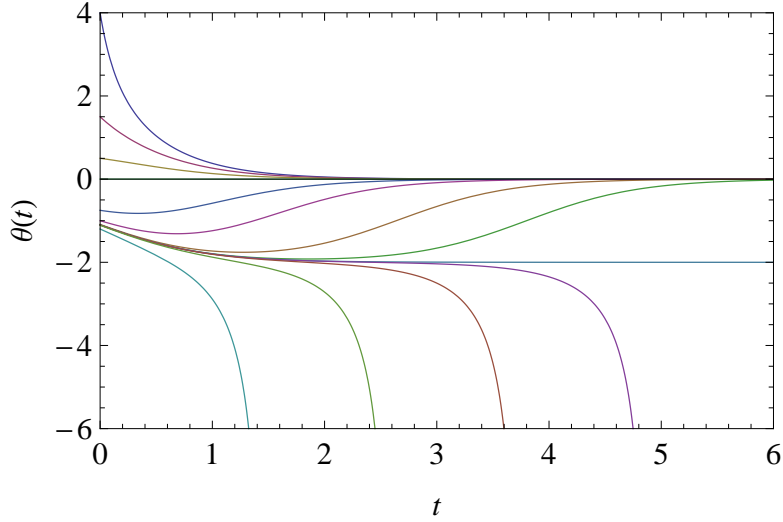}
\caption{Plots of $\theta(t)$ for different $\theta_0$. Here, $\alpha=1$, $\beta=1$ and $u_0=0.1$.}
\label{fig:terminal1}
\end{figure}

\noindent As a third example, we consider a case similar to the
one discussed just above, 
but with the particle subject to a 
drag force quadratic in velocity. Here, the equation of motion becomes
\begin{equation}
f=\frac{du}{dt}=\frac{d^2x}{dt^2}=\alpha-\beta u^2
\label{eq:terminal1}
\end{equation}
where $\alpha$ is the constant driving force per unit mass (here $\alpha=g$) 
on the particle. The equation for the expansion turns out to be
\begin{equation}
\frac{d\theta}{dt}+\theta^2+2\beta u \theta=0
\label{eq:terminal2}
\end{equation}
Unlike the previous two examples, here the speed $u$ appears in the
equation for $\theta$. This is reminiscent of the equation for the
expansion in Riemannian geometry and Relativity (the Raychaudhuri equation)
where also the velocity appears in the equation for $\theta$.
The solutions of the above set of equations are given by
\begin{equation}
x(t)=x_0+\frac{1}{\beta}\log\left[\frac{\cosh[\sqrt{\alpha\beta}t+\tanh^{-1}(\sqrt{\frac{\beta}{\alpha}}u_0)]}{\cosh[\tanh^{-1}(\sqrt{\frac{\beta}{\alpha}}u_0)]}\right]
\label{eq:terminal3}
\end{equation}
\begin{equation}
u(t)=\sqrt{\frac{\alpha}{\beta}}\tanh[\sqrt{\alpha\beta}t+\tanh^{-1}(\sqrt{\frac{\beta}{\alpha}}u_0)]
\label{eq:terminal4}
\end{equation}
\begin{equation}
\theta(t)=\frac{\sqrt{\alpha\beta}\sech^2[\sqrt{\alpha\beta}t+\tanh^{-1}(\sqrt{\frac{\beta}{\alpha}}u_0)]}{C+\tanh[\sqrt{\alpha\beta}t+\tanh^{-1}(\sqrt{\frac{\beta}{\alpha}}u_0)]}
\label{eq:terminal5}
\end{equation}
where 
$C=\frac{\sqrt{\alpha\beta}}{\theta_0}\sech^2[\tanh^{-1}(\sqrt{\frac{\beta}{\alpha}}u_0)]-\sqrt{\frac{\beta}{\alpha}}u_0$. Here, $x_0$, $u_0$ and $\theta_0$ 
are, respectively, the initial position, velocity and expansion. If the initial expansion is positive ($\theta_0>0$), the denominator of Eqn. (\ref{eq:terminal5}) and hence the expansion $\theta(t)$ always remains positive. In the limit $t\rightarrow\infty$, $\theta(t)$ goes to $0$. For $\theta_0=0$, $\theta(t)$ 
remains zero for all times since it is a fixed point of Eqn. (\ref{eq:terminal2}). 
For $\theta_{0}<0$, meeting of trajectories may take place. 
Now, as $t\to\infty$, the $\tanh$ function in the denominator of Eqn. (\ref{eq:terminal5}) goes to $1$. 
Therefore, for the trajectories to meet in finite time, one must have $C>-1$. The case $C=-1$ gives the initial critical value 
($\theta_c$) of the expansion for which meeting takes place in infinite time, 
i.e., above this critical value no intersection of the trajectories take place. Setting $C=-1$, we get the initial critical expansion which turns out to be 
$\theta_c=-\sqrt{\alpha\beta}(1+\sqrt{\beta/\alpha}u_0)$. 
For $\theta_0=\theta_c$, $\theta(t)$ approaches the limiting value 
$\underset{t\to\infty}{\lim} \theta(t)=-2\sqrt{\alpha\beta}$ as $t\rightarrow\infty$. These features are reflected in Fig.~\ref{fig:terminal1} which
shows the plot of $\theta(t)$ for different $\theta_0$.

\begin{figure}[ht]
\centering
\includegraphics[scale=0.88]{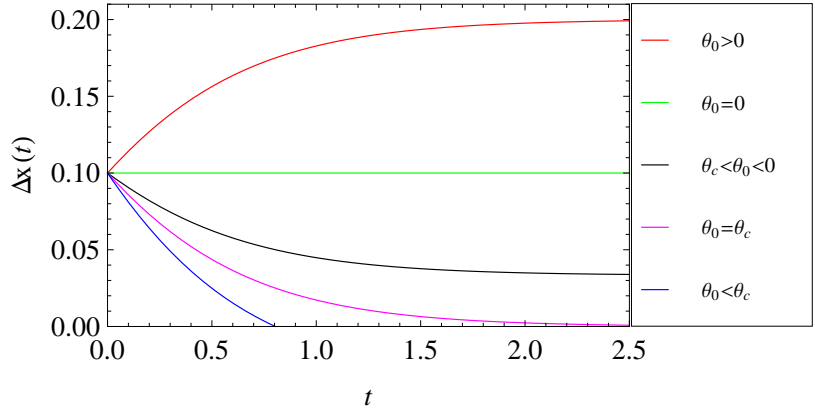}
\caption{Plots of the separation $\Delta x(t)$ between two trajectories for different $\theta_0$.}
\label{fig:terminal2}
\end{figure}

\noindent As in the previous examples, we take two trajectories and plot the difference $\Delta x(t)$ as a function of time. When $\Delta x(t_f)=0$  the
trajectories meet and  $\theta(t_f)\to -\infty$. Fig.~\ref{fig:terminal2} shows 
the variation of the separation between the two trajectories for different initial expansion $\theta_{0}$.

\noindent The three examples discussed above illustrate
the basic theme of this article for the simplest scenario, i.e. in one space dimension. The equations for
$\theta(t)$ in these three examples have important differences in their
form as well as in the solutions-- a fact which has been
highlighted above. We will now turn to two space dimensions and explore
the novelties that arise.  

\section{Two dimensions}
\label{sec:two_dimension}

\subsection{Formalism}
\noindent In more than one dimension, the formalism is more complex and involved. Let us first work things out in the case of general ($>1$) dimensions
after which we specialise to two dimensions.
The generalisation of the Eqn. (\ref{eq:1d-theta-gen}) for arbitrary dimensions turns out to be
\begin{equation}
u^k\partial_k(\partial_j u^i)=\partial_j f^i-(\partial_j u^k)(\partial_k u^i)
\label{eq:ESReuclid1}
\end{equation}
where, $f^i=u^k\partial_k u^i=\frac{du^i}{dt}$ is the $i$'th component of the force per unit mass, i.e., acceleration of the particle. In $n$-dimensional 
Euclidean space, the tensor $B_{ij}=\partial_j u_i$ can be decomposed into 
its trace part $\theta$ (expansion), symmetric traceless part $\sigma_{ij}$ (shear) and antisymmetric part $\omega_{ij}$ (rotation) as follows:
\begin{equation}
B_{ij}=\partial_j u_i=\frac{1}{n}\delta_{ij}\theta+\sigma_{ij}+\omega_{ij}
\label{eq:decomposegeneral}
\end{equation}
where $\theta=B^i_{\;i}$, $\sigma_{ij}=\frac{1}{2}(B_{ij}+B_{ji})-\frac{1}{n}\delta_{ij}\theta$ and $\omega_{ij}=\frac{1}{2}(B_{ij}-B_{ji})$. The Eqn. (\ref{eq:ESReuclid1}) can be rewritten as
\begin{equation}
u^k\partial_k B^i_{\;j}=\partial_j f^i - B^i_{\;k} B^k_{\;j}
\label{eq:ESReuclid2}
\end{equation}
One can easily extract the evolution equations for 
$\theta$, $\sigma_{ij}$ and $\omega_{ij}$ from (\ref{eq:decomposegeneral}) 
and (\ref{eq:ESReuclid2}). 
It should be noted that, for a given velocity vector $u^i$, one can calculate the expansion, rotation and shear though that does not accommodate arbitrary initial conditions on the ESR variables of the family of trajectories. 
The Eqn. (\ref{eq:ESReuclid2}), on the other hand, gives a set of evolution equations for the ESR variables and hence accommodates arbitrary initial conditions. Therefore, 
Eqn. (\ref{eq:ESReuclid2}) is more general.

\noindent Let us now specialise to two dimensions. 
In two dimensions, expansion, shear and rotation are nonzero, in general. 
We decompose the tensor $B^i_j=\partial_j u^i$ in the following way \cite{poisson}.
\begin{equation}
B^i_j=\partial_j u^i=\left(\begin{matrix}\frac{1}{2}\theta & 0 \\ 0 & \frac{1}{2}\theta \end{matrix}\right) + \left(\begin{matrix} \sigma_{+} & \sigma_{\times} \\ \sigma_{\times} & -\sigma_{+} \end{matrix}\right) + \left(\begin{matrix}
0 & \omega \\ -\omega & 0
\end{matrix}\right)
\label{eq:2dESRdecompose}
\end{equation}
From Eqn. (\ref{eq:ESReuclid2}) and Eqn. (\ref{eq:2dESRdecompose}), we 
obtain the following equations for $\theta$, $\sigma_{+}$, $\sigma_{\times}$ and $\omega$ :
\begin{equation}
\frac{d\theta}{dt}+\frac{1}{2}\theta^2+2\left(\sigma_{+}^2+\sigma_{\times}^2-\omega^2\right)=\left(\frac{\partial f_x}{\partial x}+\frac{\partial f_y}{\partial y}\right)
\label{eq:2dESR-eqn1}
\end{equation}
\begin{equation}
\frac{d\sigma_{+}}{dt}+\theta\sigma_{+}=\frac{1}{2}\left(\frac{\partial f_x}{\partial x}-\frac{\partial f_y}{\partial y}\right)
\label{eq:2dESR-eqn2}
\end{equation}
\begin{equation}
\frac{d\sigma_{\times}}{dt}+\theta\sigma_{\times}=\frac{1}{2}\left(\frac{\partial f_x}{\partial y}+\frac{\partial f_y}{\partial x}\right)
\label{eq:2dESR-eqn3}
\end{equation}
\begin{equation}
\frac{d\omega}{dt}+\theta\omega=\frac{1}{2}\left(\frac{\partial f_x}{\partial y}-\frac{\partial f_y}{\partial x}\right)
\label{eq:2dESR-eqn4}
\end{equation}
where $f_x$ and $f_y$ are the $x$ and $y$ component of the force per unit mass, i.e., acceleration acting on the particle. It should be noted that the force on the particle may depend on velocity; in that case, the gradient of the force will contain gradient of the velocity. One should replace this gradient of the velocity coming out of the gradient of force by the ESR variables.

\noindent Similar to the one dimensional case, one may try to find out (without
solving the above equations) whether there
exists a real, finite time $t=t_f$ when a family of trajectories will meet.
Of course, $t_f$ can surely be found by solving these equations for a
given system. The important point to note is that $t_f$ will now depend
on the initial values $\theta_0$, $\sigma_{+0}$, $\sigma_{\times 0}$ and 
$\omega_0$. We will illustrate this approach towards obtaining $t_f$ for a 
specific example, later.
 
\begin{figure}[ht]
\centering
\includegraphics[scale=0.9]{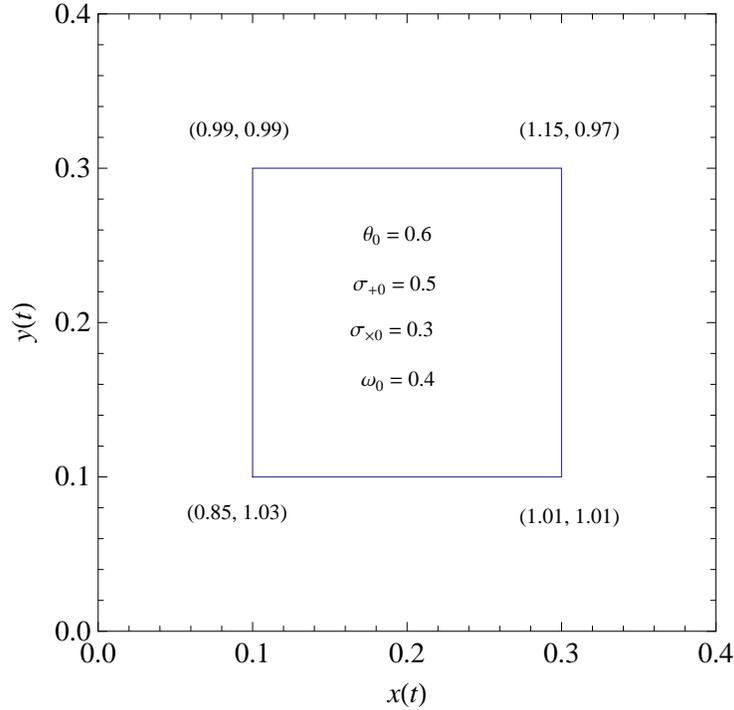}
\caption{Figure showing initial configuration of the family of
trajectories for $|\Delta x_{i0}|=0.1$ and $|\Delta y_{i0}|=0.1$. The velocity of the central trajectory is $(u_{x0},u_{y0})=(1.0,1.0)$. We consider bottom left corner as first trajectory and move counter-clockwise while numbering the other corners.}
\label{fig:squar}
\end{figure}

\noindent 
Let us now try to visualise the evolution of the ESR variables. 
We consider four trajectories which start at $t=0$, from the four corners of a square in the $xy$-plane with the 
center of the square as the starting point of the central trajectory. 
The initial position and velocity at the center of the square are $(x_0,y_0)$ and $(u_{x0},u_{y0})$, respectively. We denote the initial positions and
velocities of the four corners as $(x_{i0}, y_{i0})$ and $(u_{xi0},u_{yi0})$ where
$i=1,2,3,4$ runs anti-clockwise from the bottom left corner. The velocity $(u_{xi0},
u_{yi0})$ can be written as a Taylor series about $(u_{x0},u_{y0})$ as
\begin{eqnarray}
u_{xi0} (x_{i0}, y_{i0}) = u_{x0} (x_0,y_0) + \frac{\partial u_x}{\partial x}\Big\vert_{(x_0, y_0)} \Delta x_{i0} +  \frac{\partial u_x}{\partial y}\Big\vert_{(x_0, y_0)} \Delta y_{i0}
\label{eq:taylor_expansion_1} \\ 
u_{yi0} (x_{i0}, y_{i0}) = u_{y0} (x_0,y_0) + \frac{\partial u_y}{\partial x}\Big\vert_{(x_0, y_0)} \Delta x_{i0} +  \frac{\partial u_y}{\partial y}\Big\vert_{(x_0, y_0)} \Delta y_{i0} 
\label{eq:taylor_expansion_2}
\end{eqnarray} 
where, $\Delta x_{i0}=x_{i0}-x_0$ and $\Delta y_{i0}=y_{i0}-y_0$ are the initial separation between the central trajectory and one of the trajectories which start from the corners of the square. Therefore, knowing the four derivatives and the four quantities denoting the
position and velocity of the central trajectory, we can find the velocities at the
corners in terms of $\Delta x_{i0}$ and $\Delta y_{i0}$. Using Eqn. (\ref{eq:2dESRdecompose}), the
derivatives can be re-expressed in terms of $\theta_0$, $\sigma_{+0}$, 
$\sigma_{\times 0}$ and $\omega_0$. In terms of these quantities, the last two equations can be rewritten as
\begin{equation}
\left(\begin{matrix} u_{xi0} \\ u_{yi0} \end{matrix}\right)=\left(\begin{matrix} u_{x0} \\ u_{y0} \end{matrix}\right)+\left(\begin{matrix} \frac{\theta_0}{2}+\sigma_{+0} & & \sigma_{\times 0}+\omega_0 \\ \sigma_{\times 0}-\omega_0 & & \frac{\theta_0}{2}-\sigma_{+0} \end{matrix}\right)\left(\begin{matrix} \Delta x_{i0} \\ \Delta y_{i0} \end{matrix}\right)
\end{equation}
Therefore, specifying 
$\theta_0$, $\sigma_{+0}$,
$\sigma_{\times 0}$, $\omega_0$, as well as the values of $(x_0,y_0)$,
$(u_{x0}, u_{y0})$, $\Delta x_{i0}$ and $\Delta y_{i0}$, one can obtain the initial positions and velocities at the four corners of the square. An example illustrating 
this scheme is shown in Fig.~\ref{fig:squar}.

\noindent Once all initial values are fixed, the evolution of the
square can be found by following the set of trajectories. The 
geometric shape of the square, at different time instances
along the family of trajectories, will carry  the information
about the time evolution and interpretation of the kinematic quantities 
$\theta$, $\sigma_+$,
$\sigma_\times$ and $\omega$. We shall illustrate this aspect in all
the two-dimensional examples discussed below.

\noindent The geometric deformation of the square may also be understood from deviation equation. Let $\eta^i$ be the deviation vector of a trajectory from the central one. Here, $\eta^i=(\eta^x,\eta^y)=(\Delta x(t),\Delta y(t))$ at any time $t$. Since $\eta^i$ is the separation between the two trajectories, the time rate of change of $\eta^i$ is the difference in velocity between them at any time $t$. Although the Eqns. (\ref{eq:taylor_expansion_1}-\ref{eq:taylor_expansion_2}) are written for $t=0$, it is true for any time $t$. Therefore, using (\ref{eq:taylor_expansion_1}-\ref{eq:taylor_expansion_2}) at any time $t$, one obtain the deviation equations
\begin{eqnarray*}
\frac{d\eta^x}{dt}=u_{xi} (x_{i}, y_{i}) - u_{x} (x,y) = \frac{\partial u_x}{\partial x}\Big\vert_{(x, y)} \eta^x +  \frac{\partial u_x}{\partial y}\Big\vert_{(x, y)} \eta^y \\ 
\frac{d\eta^y}{dt}=u_{yi} (x_{i}, y_{i}) - u_{y} (x,y) = \frac{\partial u_y}{\partial x}\Big\vert_{(x, y)} \eta^x +  \frac{\partial u_y}{\partial y}\Big\vert_{(x, y)} \eta^y
\end{eqnarray*}
where, $(x,y)$ and $(x_i,y_i)$ are respectively the positions of central and any one of the trajectories at any time $t$. In tensor notation, the preceeding two equations can be written as
\begin{equation}
u^j \partial_j \eta^i=\frac{d\eta^i}{dt}=\partial_j u^i \eta^j=B^i_{\;j} \eta^j
\label{eq:deviation_equation}
\end{equation}
Using this equation, one obtain the following
\begin{equation}
{\cal L}_u \eta^i=u^j \partial_j \eta^i - \eta^j \partial_j u^i=0
\end{equation}
i.e. the Lie derivative of the deviation vector along the central trajectory vanishes.

\subsection{Examples}

\subsubsection{Projectile motion}
\noindent The equations of motion for a projectile in the $xy$-plane (with $x$-axis horizontal and $y$-axis vertical) are given by
\begin{equation}
f_x=\frac{du_x}{dt}=\frac{d^2 x}{dt^2}=0
\label{eq:projectile-force1}
\end{equation}
\begin{equation}
f_y=\frac{du_y}{dt}=\frac{d^2 y}{dt^2}=-g
\label{eq:projectile-force2}
\end{equation}
where $g$ is acceleration due to gravity. The equations for the 
ESR variables becomes
\begin{equation}
\frac{d\theta}{dt}+\frac{1}{2}\theta^2+2\left(\sigma_{+}^2+\sigma_{\times}^2-\omega^2\right)=0
\label{eq:projectile-ESR1}
\end{equation}
\begin{equation}
\frac{d\sigma_{+}}{dt}+\theta\sigma_{+}=0
\label{eq:projectile-ESR2}
\end{equation}
\begin{equation}
\frac{d\sigma_{\times}}{dt}+\theta\sigma_{\times}=0
\label{eq:projectile-ESR3}
\end{equation}
\begin{equation}
\frac{d\omega}{dt}+\theta\omega=0
\label{eq:projectile-ESR4}
\end{equation}
The solutions to the equations of motion are given by
\begin{equation}
x(t)=x_0+u_{x0}t
\label{eq:projectile-position-x}
\end{equation}
\begin{equation}
y(t)=y_0+u_{y0}t-\frac{1}{2}gt^2
\label{eq:projectile-position-y}
\end{equation}

\noindent Before we discuss the solutions of equations for the ESR, let us
try to obtain $t_f$, as promised before, by just using the solutions of
the equations of motion. We assume $(x(t), y(t))$ as the trajectory 
for the set of initial conditions $\{x_0, y_0, u_{x0}, u_{y0}\}$. 
Let us take another trajectory $(x'(t), y'(t))$ with initial conditions
$\{x_0+\Delta x_0, y_0+\Delta y_0, u_{x0} +\frac{\partial u_{x}}{\partial x}\vert_{(x_0,y_0)}
\Delta x_0 + \frac{\partial u_{x}}{\partial y}\vert_{(x_0,y_0)} \Delta y_0, u_{y0} +\frac{\partial u_{y}}{\partial x}\vert_{(x_0,y_0)}
\Delta x_0 + \frac{\partial u_{y}}{\partial y}\vert_{(x_0,y_0)} \Delta y_0$, where
$\Delta x_0 = x'_0-x_0$, $\Delta y_0= y'_0-y_0$ and $u_{x0}= u_x(x_0,y_0)$, $u_{y0}=
u_y(x_0,y_0)$. If these trajectories meet at some $t_f$, then $x(t_f)=x'(t_f)$
and $y(t_f)=y'(t_f)$. For this example, we obtain, using the 
solutions of the equations of motion and the 
definitions of the four partial derivatives: 
$\frac{\partial u_{x}}{\partial x}\vert_{(x_0,y_0)} = \frac{\theta_0}{2}+\sigma_{+0}$,
$\frac{\partial u_{x}}{\partial y}\vert_{(x_0,y_0)} = \sigma_{\times 0}+\omega_0$,
$\frac{\partial u_{y}}{\partial x}\vert_{(x_0,y_0)} = \sigma_{\times 0} - \omega$,
$\frac{\partial u_{y}}{\partial y}\vert_{(x_0,y_0)} = \frac{\theta_0}{2}-\sigma_{+0}$,
\begin{eqnarray}
\left [1+\left (\frac{\theta_0}{2}+\sigma_{+0}\right ) t_f \right ] \Delta x_0
+\left[\left (\sigma_{\times 0}+\omega_0 \right )t_f\right] \Delta y_0 = 0 \\
\left[\left (\sigma_{\times 0}-\omega_0 \right )t_f\right] \Delta x_0 +
\left [1+\left (\frac{\theta_0}{2}-\sigma_{+0}\right ) t_f\right ] \Delta y_0 =0
\end{eqnarray}
A non-trivial solution of $(\Delta x_0,\Delta y_0)$ can be found if the coefficient
determinant of the above pair of linear equations vanishes. 
Thus we have,
\begin{equation}
 \left| \begin{array}{cc}
1+\left (\frac{\theta_0}{2}+\sigma_{+0}\right ) t_f    &  \left (\sigma_{\times 0}+\omega_0 \right )t_f \\
\left (\sigma_{\times 0}-\omega_0 \right )t_f  & 1+\left (\frac{\theta_0}{2}-\sigma_{+0}\right ) t_f \end{array} \right| =0 
\end{equation} 
This  yields a quadratic equation for $t$, the possible solutions of which are
\begin{equation}
t_f=-\frac{2}{\theta_0\pm 2\sqrt{I_0}} 
\label{eq:projectile_focusing_time}
\end{equation}
where $I_0 = \sigma_{+0}^2+\sigma_{\times 0}^2-\omega_0^2$. It is
clear that $t_f$ is real as long as $I_0\geq 0$. Also, for $t_f>0$, 
the values of $\theta_0$ and $I_0$ must be appropriately chosen. 
If $I_0<0$ then the trajectories will never meet.
We will now obtain the same results below, using exact solutions of the
equations for $\theta$, $\sigma_+$, $\sigma_\times$ and $\omega$. Also, we will see that only the focusing time $t_f$ with the minus sign in the denominator is acceptable.

\noindent In order to solve Eqns. (\ref{eq:projectile-ESR1}-\ref{eq:projectile-ESR4}), we 
introduce a quantity $I=\sigma_{+}^2+\sigma_{\times}^2-\omega^2$. Therefore, one can write down the following two equations,
\begin{equation}
\frac{d\theta}{dt}+\frac{1}2\theta^2+2I=0
\label{eq:projectile-I1}
\end{equation}
\begin{equation}
\frac{dI}{dt}+2\theta I=0
\label{eq:projectile-I2}
\end{equation}
The general solution of Eqn. (\ref{eq:projectile-I2}) is of the form $I(t)=I_0 e^{-2\int_0^t \theta(t)dt}$, where $I_0$ is the value at $t=0$. It is clear that during its evolution $I(t)$ does not change sign. Therefore, there are three sets of solutions for the three cases $I>0$, $I=0$ and $I<0$. Using this procedure, the solutions of these equations are obtained in \cite{ADG1}. The solutions are
\newline
1. $I>0$
\begin{equation}
\theta(t)=\frac{C\left(D+\frac{C}{4}t\right)}{2\left[\left(D+\frac{C}{4}t\right)^2-16\right]}
\label{eq:projectile-I1-ESR1}
\end{equation}
\begin{equation}
\left\lbrace\sigma_{+},\sigma_{\times},\omega\right\rbrace=\frac{\left\lbrace E,F,G \right\rbrace}{\left[\left(D+\frac{C}{4}t\right)^2-16\right]}
\label{eq:projectile-I1-ESR2}
\end{equation}
2. $I=0$
\begin{equation}
\theta(t)=\frac{\theta_{0}}{\left(1+\frac{\theta_{0}}{2}t\right)}
\label{eq:projectile-I2-ESR1}
\end{equation}
\begin{equation}
\left\lbrace\sigma_{+},\sigma_{\times},\omega\right\rbrace=\frac{\left\lbrace \sigma_{+0},\sigma_{\times 0},\omega_0\right\rbrace}{\left(1+\frac{\theta_{0}}{2}t\right)^2}
\label{eq:projectile-I2-ESR2}
\end{equation}
3. $I<0$
\begin{equation}
\theta(t)=\frac{C\left(D+\frac{C}{4}t\right)}{2\left[\left(D+\frac{C}{4}t\right)^2+16\right]}
\label{eq:projectile-I3-ESR1}
\end{equation}
\begin{equation}
\left\lbrace\sigma_{+},\sigma_{\times},\omega\right\rbrace=\frac{\left\lbrace E,F,G \right\rbrace}{\left[\left(D+\frac{C}{4}t\right)^2+16\right]}
\label{eq:projectile-I3-ESR2}
\end{equation}
where $\theta_0$, $\sigma_{+0}$, $\sigma_{\times 0}$ and $\omega_0$ are the initial value of the ESR variables at $t=0$. Other integration constants can be written as
\begin{equation*}
D=\frac{2\theta_0}{\sqrt{\pm I_0}}
\end{equation*}
\begin{equation*}
C=\sqrt{\pm I_0}(D^2\mp 16)
\end{equation*}
\begin{equation*}
\lbrace E,F,G\rbrace=\lbrace \sigma_{+0},\sigma_{\times 0},\omega_0 \rbrace (D^2\mp 16)
\end{equation*}
where the upper (lower) sign is for $I_0>0$ ($I_0<0$). For $I_0<0$ focusing never takes place. But, for $I_0>0$ focusing takes place only when $D<4$, i.e., $\theta_0 < 2\sqrt{I_0}$. Therefore, for $I_0>0$, there exists a critical initial rotation $\omega_c= \sqrt{\sigma_{+0}^2+\sigma_{\times 0}^2-\frac{\theta_0^2}{4}}$ above which no focusing takes place. For $I_0=0$, focusing takes place only for negative $\theta_0$. The focusing time $t_f$ obtained by setting $\theta(t_f)\to -\infty$ is given by
\begin{equation}
t_f=-\frac{2}{\theta_0- 2\sqrt{I_0}}
\end{equation}
As mentioned earlier, this $t_f$ matches with that (with the minus sign in the denominator) obtained in Eqn. (\ref{eq:projectile_focusing_time}). Clearly, for $\omega_0=\omega_c$, i.e., $\theta_0 = 2\sqrt{I_0}$ focusing time is infinite which implies that above this critical value $\omega_c$, i.e., for $\theta_0 > 2\sqrt{I_0}$, focusing never takes place.

\begin{figure}[ht]
\centering
\subfigure[$\theta_0=-0.5$, $\sigma_{+0}=0.0$, $\sigma_{\times0}=0.0$, $\omega_0=0.0$]{\includegraphics[scale=0.96]{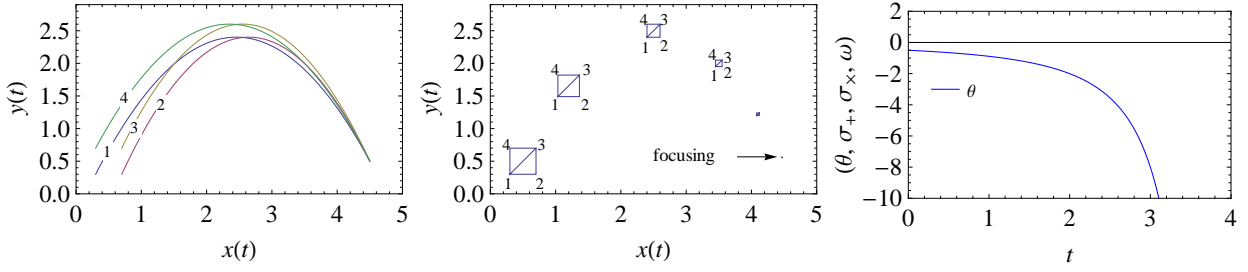}\label{fig:projectile1a}}
\subfigure[$\theta_{0}=0.25$, $\sigma_{+0}=0.4$, $\sigma_{\times0}=0.3$, $\omega_{0}=0.5$]{\includegraphics[scale=0.96]{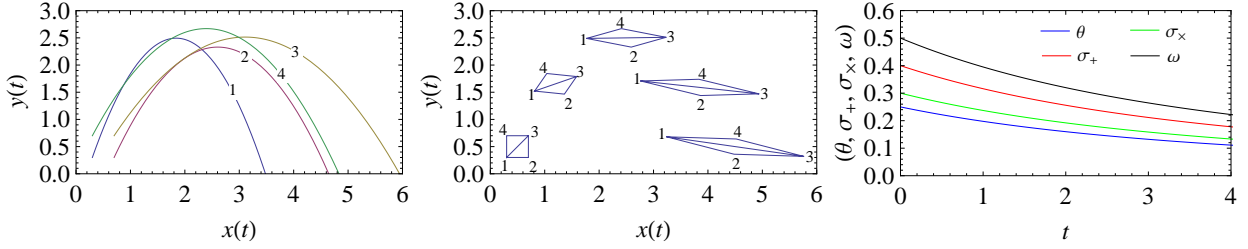}\label{fig:projectile1b}}
\subfigure[$\theta_{0}=-0.55$, $\sigma_{+0}=0.4$, $\sigma_{\times0}=0.3$, $\omega_{0}=0.5$]{\includegraphics[scale=0.96]{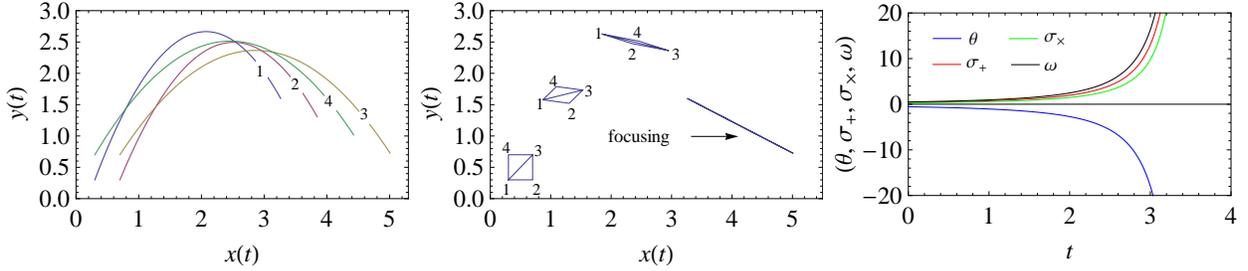}\label{fig:projectile1c}}
\caption{Plots for projectile motion for $I=0$. The initial position and velocity of the central projectile is $(0.5,0.5)$ and $(1.0,2.0)$, respectively. Here, $g=1$.}
\label{fig:projectile1}
\end{figure}

\begin{figure}[ht]
\centering
\subfigure[$\theta_{0}=-0.3$, $\sigma_{+0}=0.0$, $\sigma_{\times0}=0.1$, $\omega_{0}=0.0$]{\includegraphics[scale=0.96]{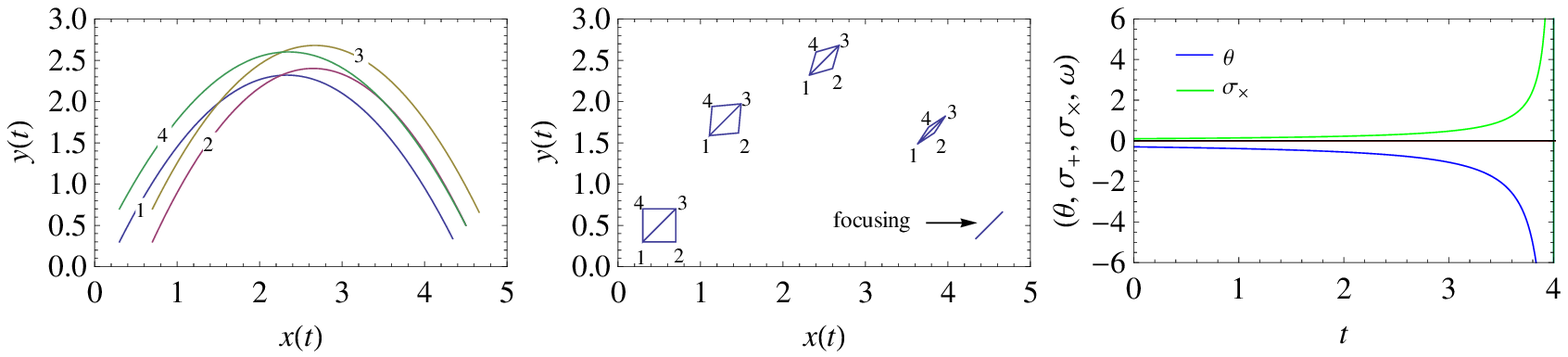}}
\subfigure[$\theta_{0}=0.1$, $\sigma_{+0}=0.3$, $\sigma_{\times0}=0.0$, $\omega_{0}=0.0$]{\includegraphics[scale=0.96]{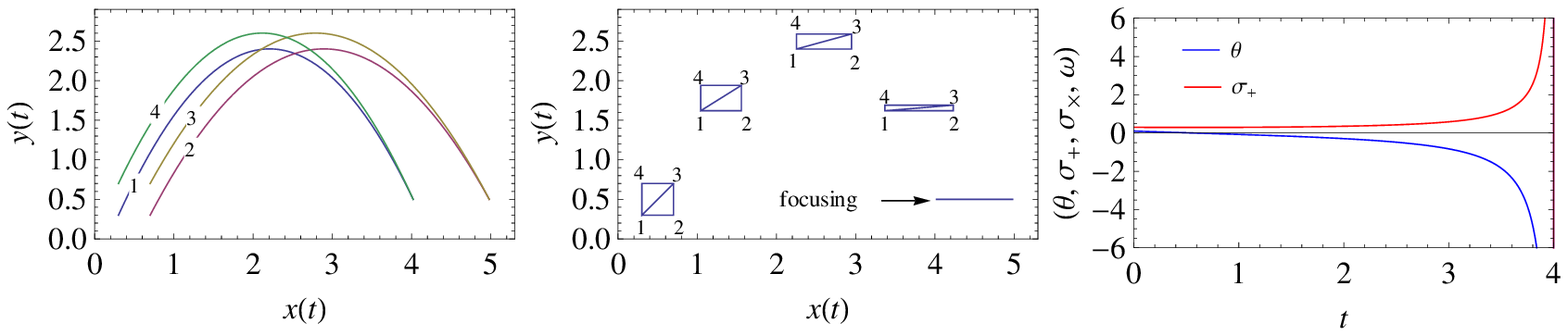}}
\subfigure[$\theta_{0}=0.3$, $\sigma_{+0}=0.4$, $\sigma_{\times0}=0.3$, $\omega_{0}=0.2$]{\includegraphics[scale=0.96]{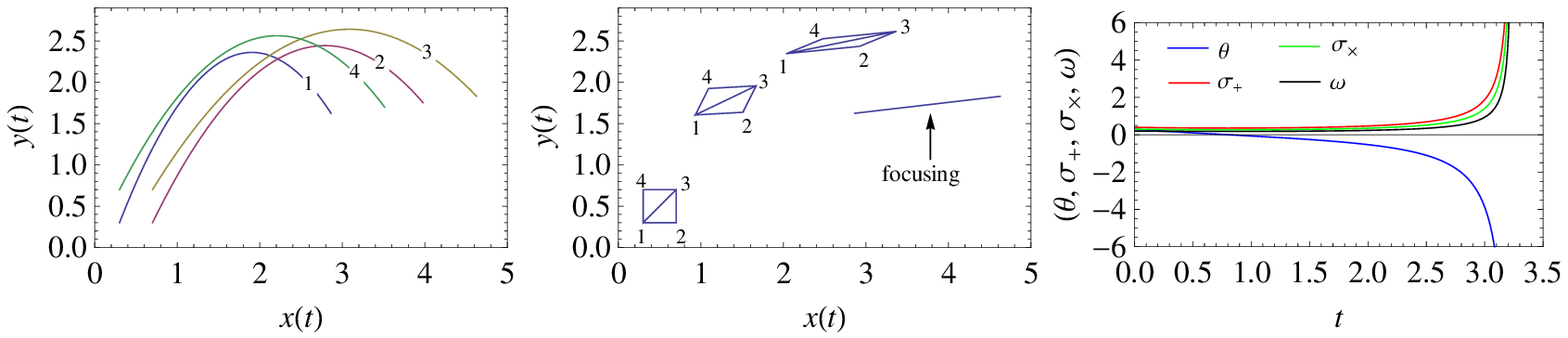}}
\caption{Plots for projectile motion for $I>0$. The initial position and velocity of the central projectile is $(0.5,0.5)$ and $(1.0,2.0)$, respectively. Here, $g=1$.}
\label{fig:projectile2}
\end{figure}

\begin{figure}[ht]
\centering
\subfigure[$\theta_{0}=-1.0$, $\sigma_{+0}=0.0$, $\sigma_{\times0}=0.0$, $\omega_{0}=0.5$]{\includegraphics[scale=0.96]{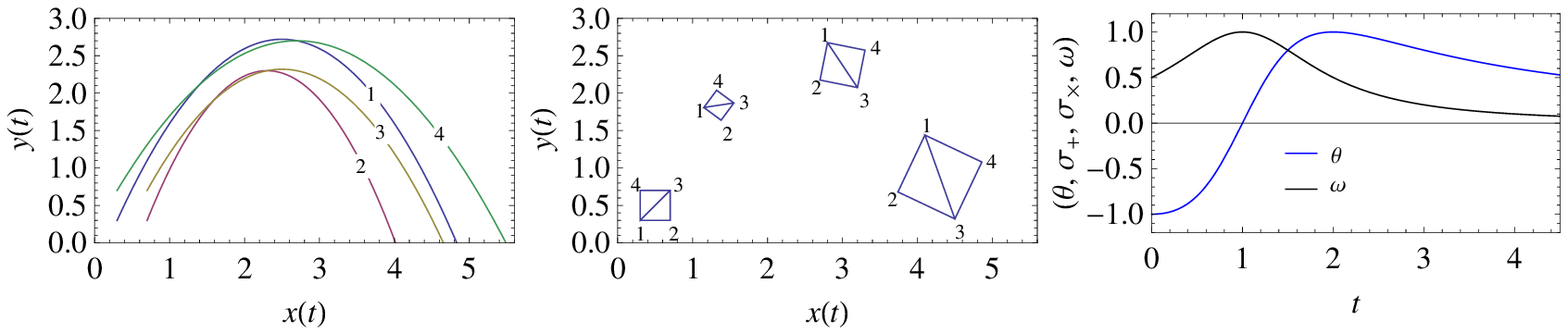}}
\subfigure[$\theta_{0}=-1.0$, $\sigma_{+0}=0.3$, $\sigma_{\times0}=0.2$, $\omega_{0}=0.5$]{\includegraphics[scale=0.96]{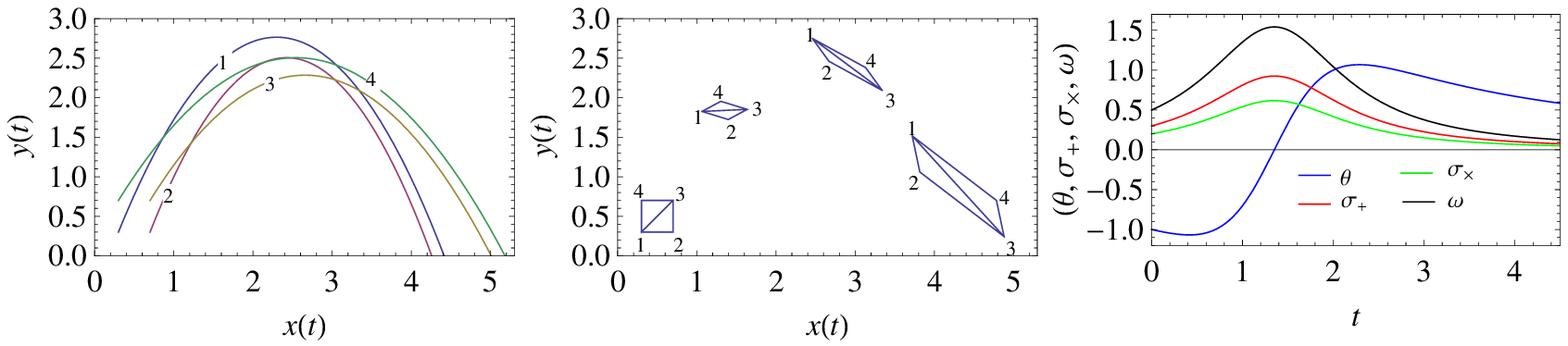}}
\caption{Plots for projectile motion for $I<0$. The initial position and velocity of the central projectile is $(0.5,0.5)$ and $(1.0,2.0)$, respectively. Here, $g=1$.}
\label{fig:projectile3}
\end{figure}

\noindent To visualise the effect of the ESR variables on the evolution of a family of trajectories, we consider four projectile 
trajectories starting out from the four corners of a square, at $t=0$. 
The initial velocities of the projectiles are calculated from the ESR variables as discussed earlier. We have plotted the four projectile trajectories
with different initial positions and velocities using Eqns. (\ref{eq:projectile-position-x}-\ref{eq:projectile-position-y}). 
These are shown in Fig.~\ref{fig:projectile1}-\ref{fig:projectile3}. In these plots, the central projectile is not shown. 
It is clear that the area of the square goes to zero whenever focusing takes place. In two dimensions, the expansion 
scalar represents the fractional rate of change of the area enclosing a family of trajectories. 
The role of expansion, shear and rotation is clear from the evolution of the square enclosing the four projectiles. From the figures, we 
notice that focusing takes place whenever $I_0\geq 0$ and $\theta_0 < 2\sqrt{I_0}$ ($\omega_0<\omega_c$) 
as discussed earlier. In Fig.~\ref{fig:projectile1b}, focusing does not take place since $\theta_0 > 2\sqrt{I_0}$, 
i.e., $\omega_0>\omega_c$. For $I_0<0$, the square shrinks initially if the initial expansion scalar is negative; 
as the evolution proceeds, the square starts expanding when expansion scalar becomes positive (Fig. \ref{fig:projectile3}). 
The vital role of initial rotation is also brought out from the figures. A sufficiently large initial rotation 
which makes $I_0<0$, or $\theta_0 > 2\sqrt{I_0}$ helps in avoiding the focusing of trajectories.

\subsubsection{Two dimensional harmonic oscillator}
\noindent For a two dimensional isotropic harmonic oscillator 
the equations of motion are given by
\begin{equation}
f_x=\frac{du_x}{dt}=\frac{d^2 x}{dt^2}=-\alpha^2 x
\label{eq:2doscillator-force1}
\end{equation}
\begin{equation}
f_y=\frac{du_y}{dt}=\frac{d^2 y}{dt^2}=-\alpha^2 y
\label{eq:2doscillator-force2}
\end{equation}
which yields 
\begin{equation}
x(t)=\sqrt{x_0^2+\frac{u_{x0}^2}{\alpha^2}}\cos(\alpha t-\phi_x).
\label{eq:2doscillator-position-x}
\end{equation}
\begin{equation}
y(t)=\sqrt{y_0^2+\frac{u_{y0}^2}{\alpha^2}}\cos(\alpha t-\phi_y)
\label{eq:2doscillator-position-y}
\end{equation}
where $\phi_x=\tan^{-1}(\frac{u_{x0}}{\alpha x_0})$ and $\phi_y=\tan^{-1}(\frac{u_{y0}}{\alpha y_0})$ are the phases. Here, $\alpha$, $(x_0,y_0)$ and $(u_{x0},u_{y0})$ are frequency of oscillation, initial position and initial velocity respectively. Before solving the ESR equations, let us find out the focusing time $t_f$ from the solution $(x(t),y(t))$ following the same prescription described in the previous example. The focusing time $t_f$ is found by solving the equation
\begin{equation}
\left| \begin{array}{cc}
\cos(\alpha t_f)+\frac{1}{\alpha}\left (\frac{\theta_0}{2}+\sigma_{+0}\right ) \sin(\alpha t_f)    &  \frac{1}{\alpha}\left (\sigma_{\times 0}+\omega_0 \right ) \sin(\alpha t_f) \\
\frac{1}{\alpha}\left (\sigma_{\times 0}-\omega_0 \right ) \sin(\alpha t_f)  & \cos(\alpha t_f)+\frac{1}{\alpha}\left (\frac{\theta_0}{2}-\sigma_{+0}\right ) \sin(\alpha t_f) \end{array} \right| =0
\end{equation}
which gives,
\begin{equation}
t_f=\frac{1}{\alpha}\tan^{-1}\left(-\frac{2\alpha}{\theta_0 \pm 2\sqrt{I_0}}\right)
\label{eq:2doscillator_fcusing_time}
\end{equation}
It is clear that one must have $I_0\geq 0$ for focusing to take place. If $I_0<0$, then focusing never takes place. 
In the following, these results are obtained from the exact solutions of the ESR equations.

\noindent The evolution equations for the ESR variables turn out to be
\begin{equation}
\frac{d\theta}{dt}+\frac{1}{2}\theta^2+2\left(\sigma_{+}^{2}+\sigma_{\times}^{2}-\omega^2\right)+2\alpha^2=0
\label{eq:2doscillator-ESR1}
\end{equation}
\begin{equation}
\frac{d\sigma_{+}}{dt}+\theta\sigma_{+}=0
\label{eq:2doscillator-ESR2}
\end{equation}
\begin{equation}
\frac{d\sigma_{\times}}{dt}+\theta\sigma_{\times}=0
\label{eq:2doscillator-ESR3}
\end{equation}
\begin{equation}
\frac{d\omega}{dt}+\theta\omega=0
\label{eq:2doscillator-ESR4}
\end{equation}
The solution of these equations are available in \cite{ADG1}. The solutions are given by
\newline
1. $I>0$
\begin{equation}
\theta(t)=\frac{2\alpha\sqrt{1+C^2}\cos(2\alpha t+D)}{C+\sqrt{1+C^2}\sin(2\alpha t+D)}
\label{eq:2doscillator-I1-ESR1}
\end{equation}
\begin{equation}
\left\lbrace\sigma_{+},\sigma_{\times},\omega\right\rbrace=\frac{\left\lbrace E,F,G \right\rbrace}{C+\sqrt{1+C^{2}}\sin(2\alpha t+D)}
\label{eq:2doscillator-I1-ESR2}
\end{equation}
where
\begin{equation*}
C=\frac{\sqrt{I_0}}{2\alpha}\left[-1+\frac{\theta_0^2}{4I_0}+\frac{\alpha^2}{I_0}\right]
\end{equation*}
\begin{equation*}
D=\tan^{-1}\left[\frac{2\sqrt{I_0}}{\theta_0}\left(\frac{\alpha}{\sqrt{I_0}}-C\right)\right]
\end{equation*}
\begin{equation*}
\lbrace E,F,G \rbrace=\frac{\alpha}{\sqrt{I_0}}\lbrace \sigma_{+0},\sigma_{\times 0},\omega_0 \rbrace
\end{equation*}
2. $I=0$
\begin{equation}
\theta(t)=2\alpha\tan[\alpha(C-t)]
\label{eq:2doscillator-I2-ESR1}
\end{equation}
\begin{equation}
\left\lbrace\sigma_{+},\sigma_{\times},\omega\right\rbrace=\left\lbrace E,F,G\right\rbrace\sec^2[\alpha(C-t)]
\label{eq:2doscillator-I2-ESR2}
\end{equation}
where
\begin{equation*}
C=\frac{1}{\alpha}\tan^{-1}\left(\frac{\theta_0}{2\alpha}\right)
\end{equation*}
\begin{equation*}
\lbrace E,F,G \rbrace=\frac{\lbrace \sigma_{+0},\sigma_{\times 0},\omega_0 \rbrace}{1+\left(\frac{\theta_0}{2\alpha}\right)^2}
\end{equation*}
3. $I<0$
\begin{equation}
\theta(t)=\frac{2\alpha C\cos(2\alpha t+D)}{\sqrt{1+C^2}+C\sin(2\alpha t+D)}
\label{eq:2doscillator-I3-ESR1}
\end{equation}
\begin{equation}
\left\lbrace\sigma_{+},\sigma_{\times},\omega\right\rbrace=\frac{\left\lbrace E,F,G \right\rbrace}{\sqrt{1+C^2}+C\sin(2\alpha t+D)}
\label{eq:2doscillator-I3-ESR2}
\end{equation}
where
\begin{equation*}
C=\sqrt{-\frac{I_0}{4\alpha^2}\left(1-\frac{\theta_0^2}{4I_0}-\frac{\alpha^2}{I_0}\right)^2-1}
\end{equation*}
\begin{equation*}
D=\tan^{-1}\left[\frac{2\sqrt{-I_0}}{\theta_0}\left(\frac{\alpha}{\sqrt{-I_0}}-\sqrt{1+C^2}\right)\right]
\end{equation*}
\begin{equation*}
\lbrace E,F,G \rbrace=\frac{\alpha}{\sqrt{-I_0}}\lbrace \sigma_{+0},\sigma_{\times 0},\omega_0 \rbrace
\end{equation*}
From the solutions, one can note that, for $I\geq 0$, focusing always takes place in finite time. For the case $I<0$, focusing never takes place because the denominator never becomes zero; the evolution of the ESR variables are oscillatory in nature. From (\ref{eq:2doscillator-I1-ESR1}), it is clear that, for the divergence of $\theta(t)$ at some time $t_f$, one must have $C+\sqrt{1+C^2}\sin(2\alpha t_f+D)=0$. One can show that $\theta(t_f) \to -\infty$ and the focusing time $t_f$ thus obtained is the same as that obtained in Eqn. (\ref{eq:2doscillator_fcusing_time}).

\begin{figure}[ht]
\centering
\subfigure[$\theta_{0}=3.0$, $\sigma_{+0}=0.0$, $\sigma_{\times0}=0.0$, $\omega_{0}=0.0$]{\includegraphics[scale=0.96]{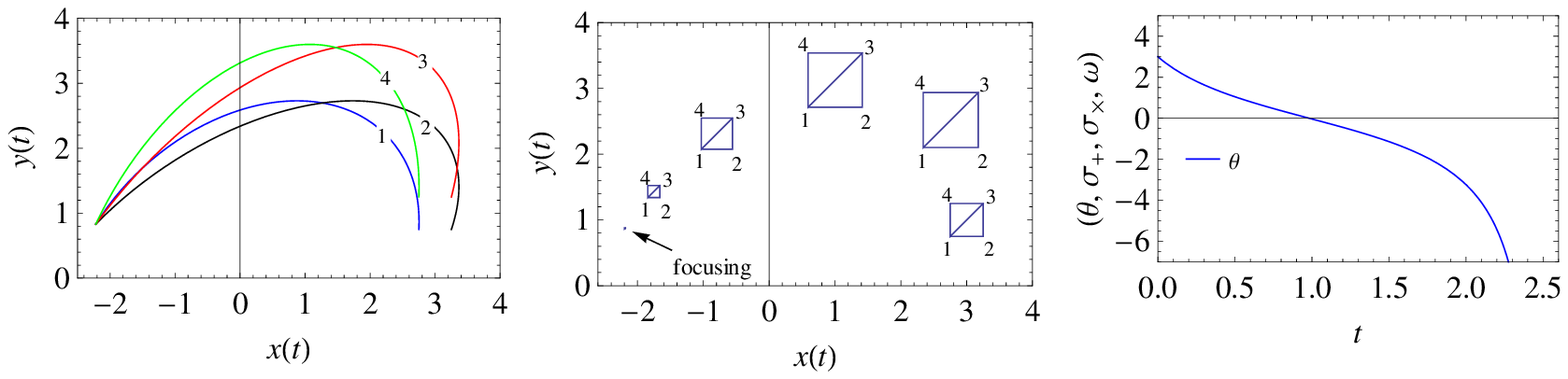}\label{fig:2doscillator1a}}
\subfigure[$\theta_{0}=3.0$, $\sigma_{+0}=0.4$, $\sigma_{\times0}=0.3$, $\omega_{0}=0.5$]{\includegraphics[scale=0.96]{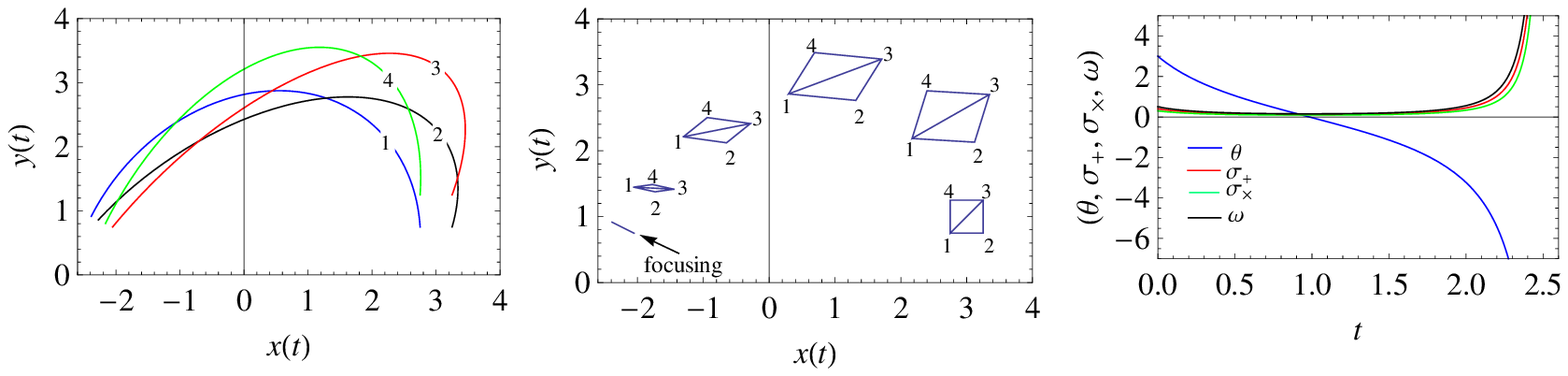}}
\caption{Plots for two dimensional oscillator for $I=0$. The initial position and velocity of the central oscillator is $(3.0, 1.0)$ and $(0.5, 3.0)$, respectively. Here, $\alpha=1$.}
\label{fig:2doscillator1}
\end{figure}

\begin{figure}[ht]
\centering
\subfigure[$\theta_{0}=3.0$, $\sigma_{+0}=0.6$, $\sigma_{\times0}=0.0$, $\omega_{0}=0.0$]{\includegraphics[scale=0.93]{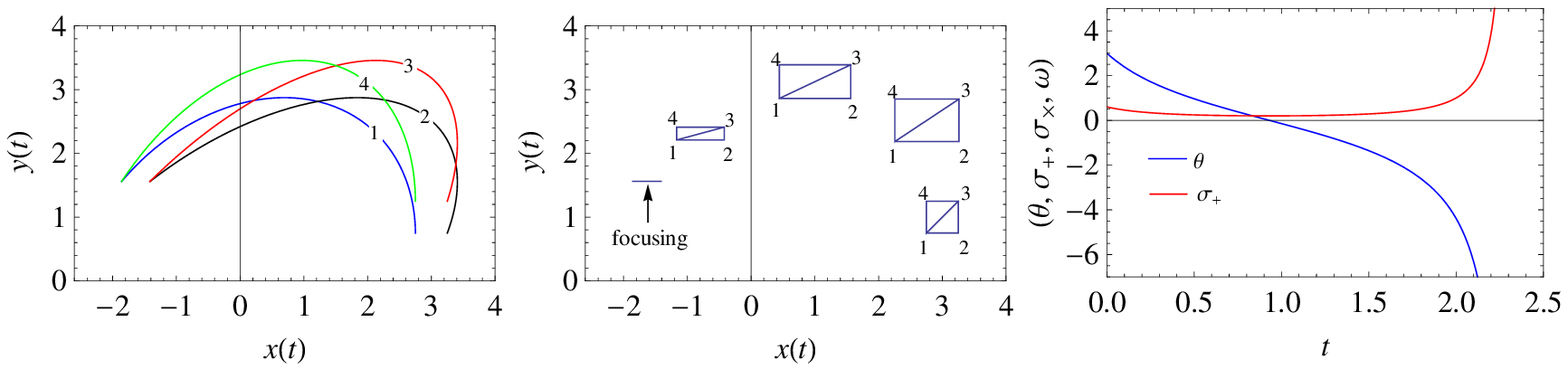}\label{fig:2doscillator2a}}
\subfigure[$\theta_{0}=3.0$, $\sigma_{+0}=0.0$, $\sigma_{\times0}=0.3$, $\omega_{0}=0.0$]{\includegraphics[scale=0.93]{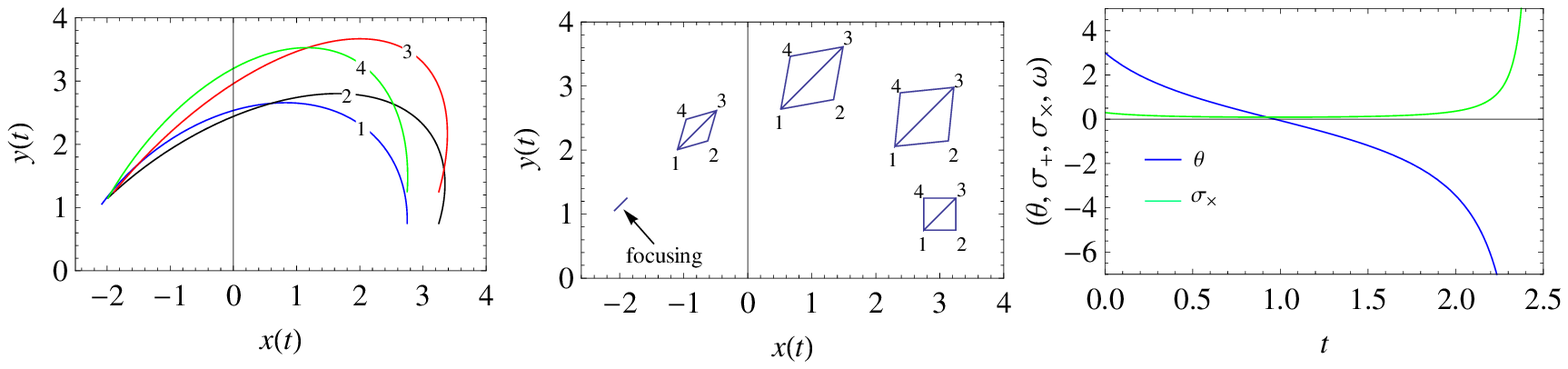}\label{fig:2doscillator2b}}
\subfigure[$\theta_{0}=3.0$, $\sigma_{+0}=0.7$, $\sigma_{\times0}=0.3$, $\omega_{0}=0.5$]{\includegraphics[scale=0.93]{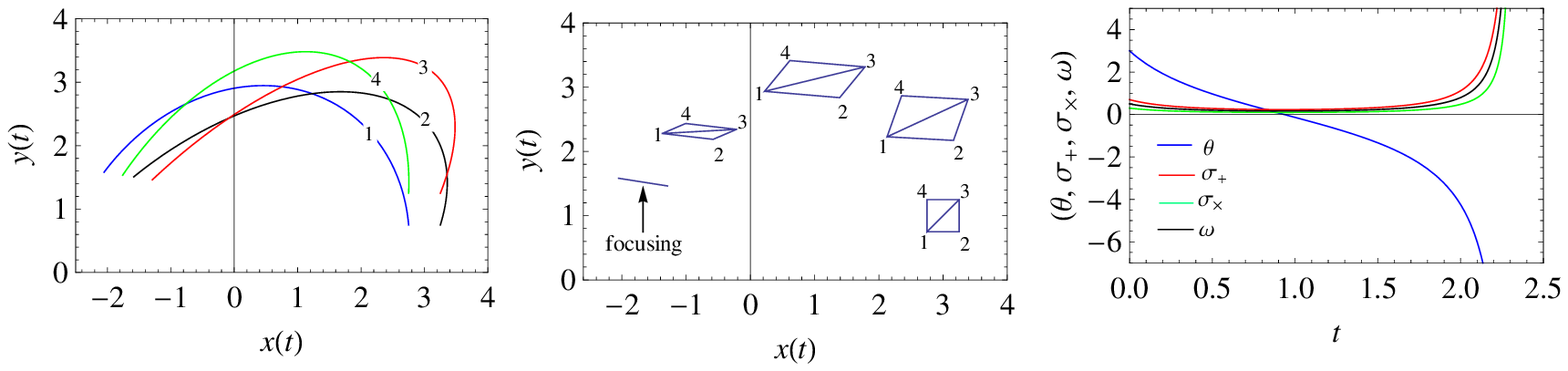}}
\caption{Plots for two dimensional oscillator for $I>0$. The initial position and velocity of the central oscillator is $(3.0, 1.0)$ and $(0.5, 3.0)$, respectively. Here, $\alpha=1$.}
\label{fig:2doscillator2}
\end{figure}

\begin{figure}[ht]
\centering
\subfigure[$\theta_{0}=3.0$, $\sigma_{+0}=0.0$, $\sigma_{\times0}=0.0$, $\omega_{0}=1.0$]{\includegraphics[scale=0.96]{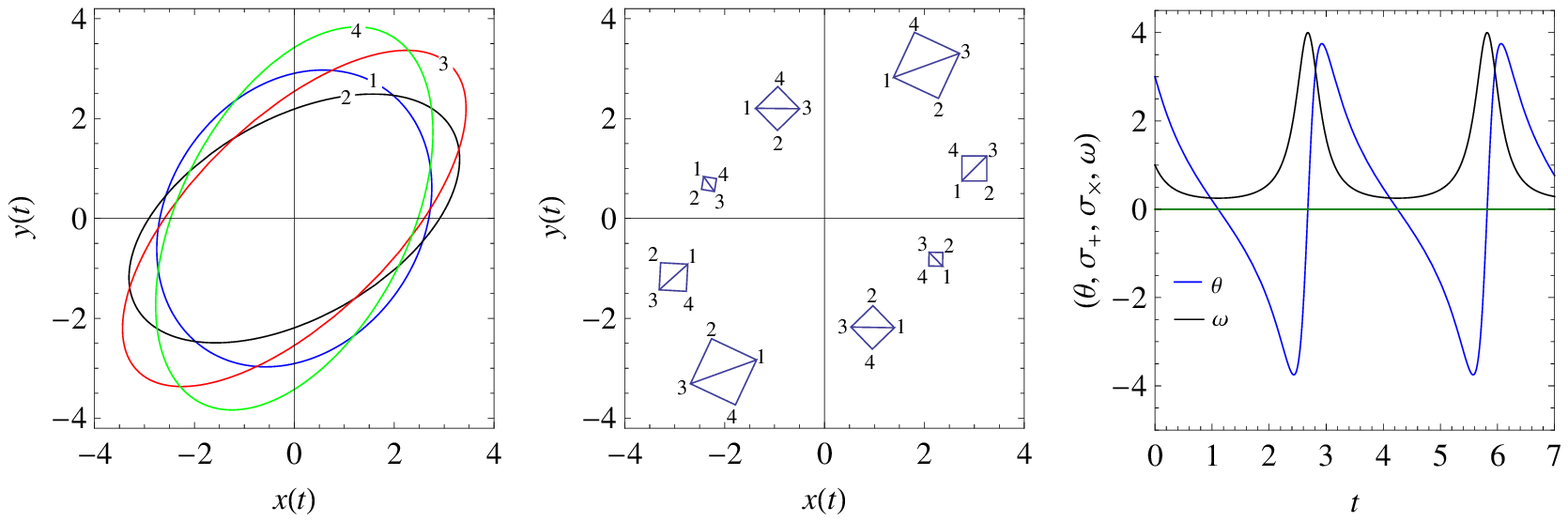}}
\subfigure[$\theta_{0}=3.0$, $\sigma_{+0}=0.5$, $\sigma_{\times0}=0.4$, $\omega_{0}=1.0$]{\includegraphics[scale=0.96]{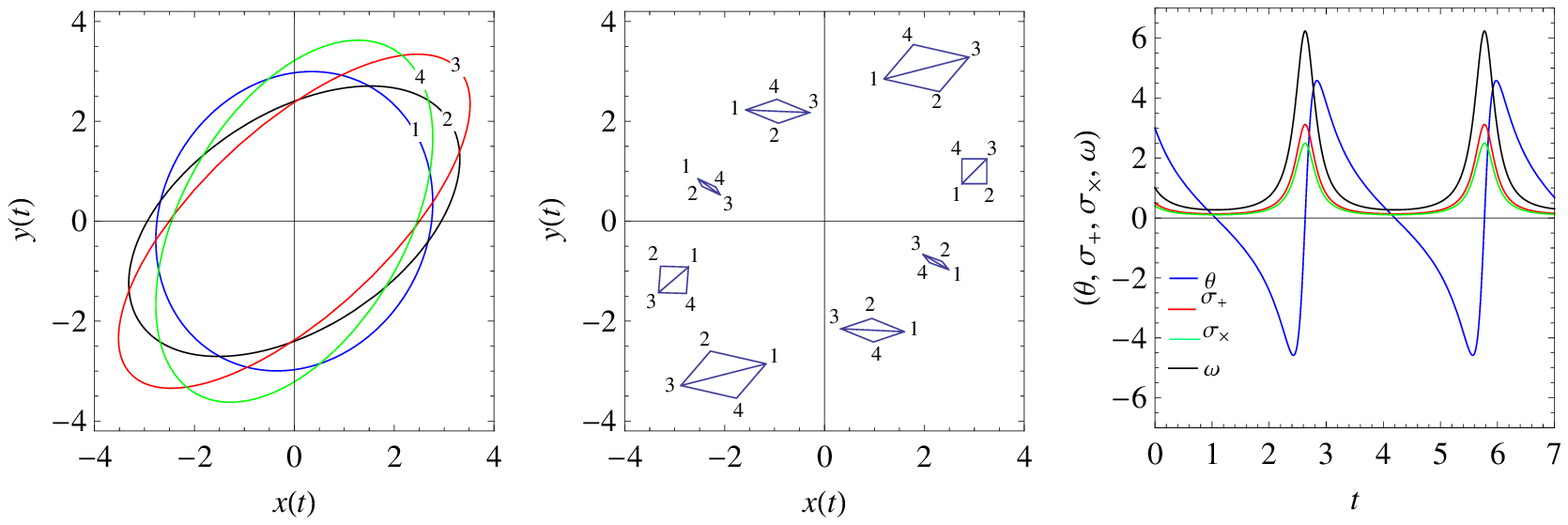}}
\caption{Plots for two dimensional oscillator for $I<0$. The initial position and velocity of the central oscillator is $(3.0, 1.0)$ and $(0.5, 3.0)$, respectively. Here, $\alpha=1$.}
\label{fig:2doscillator3}
\end{figure}

\noindent Similar to the case of projectile motion, here too we consider 
four oscillator trajectories starting from four corners of a square. 
Their initial velocities are calculated from the ESR variables 
as before. Figs.~\ref{fig:2doscillator1}-\ref{fig:2doscillator3} demonstrate
the behaviour of the family of trajectories. The role of shear and rotation,
as well as expansion, is clear from the evolution of the area enclosing
the family of trajectories. From the figures, one notices that, depending on the values of $\theta_0$, $\sigma_{+0}$, $\sigma_{\times 0}$ and $\omega_0$,
the family of trajectories can exhibit isotropic expansion or contraction (Fig.~\ref{fig:2doscillator1a}), lateral shear (Fig.~\ref{fig:2doscillator2a}), 
pure shear (Fig.~\ref{fig:2doscillator2b}) and rotation of the square, respectively. Meeting of trajectories 
takes place for $I_0\geq0$. For $I_0<0$, the evolution of the ESR variables, and hence the change in area 
of the square, shows an oscillatory nature (Fig.~\ref{fig:2doscillator3}). As the evolution proceeds, the 
square shrinks when $\theta$ is negative and expands when $\theta$ is positive; but the area of the 
square never goes to zero, which implies that focusing does not occur, 
i.e., $\theta$ never diverges 
to negative infinity. Once again, one can notice the vital role of rotation in focusing/defocussing. 
Sufficiently large initial rotation makes $I_0<0$, which prevents the meeting of trajectories.

\subsubsection{Charged particle in electromagnetic (EM) field}
\noindent Our final example involves a particle of mass $m$ and charge $q$ 
moving in crossed, constant electric and magnetic fields where 
the magnetic field is chosen to be along the $z$-axis and the electric field 
is in the $xy$-plane. We also assume that the particle is restricted to 
move in the $xy$-plane. Therefore, the problem is effectively two dimensional. The equations of motion are
\begin{equation}
f_x=\frac{du_x}{dt}=\frac{d^2x}{dt^2}=\frac{q}{m}E_x+\frac{qB}{m}u_y
\label{eq:EMfield-force1}
\end{equation}
\begin{equation}
f_y=\frac{du_y}{dt}=\frac{d^2y}{dt^2}=\frac{q}{m}E_y-\frac{qB}{m}u_x
\label{eq:EMfield-force2}
\end{equation}
where $B$, $E_x$ and $E_y$ are constant magnitudes of the magnetic field, $x$-component of the electric field and $y$-component of the electric field, 
respectively. The solutions of these equations of motion are given by
\begin{equation}
x(t)=x_0+\frac{E_y}{B}t+\left(\frac{u_{x0}}{\alpha}-\frac{E_y}{\alpha B}\right)\sin(\alpha t)+\left(\frac{u_{y0}}{\alpha}+\frac{E_x}{\alpha B}\right)[1-\cos(\alpha t)]
\label{eq:EMfield-position-x}
\end{equation}
\begin{equation}
y(t)=y_0-\frac{E_x}{B}t-\left(\frac{u_{x0}}{\alpha}-\frac{E_y}{\alpha B}\right)[1-\cos(\alpha t)]+\left(\frac{u_{y0}}{\alpha}+\frac{E_x}{\alpha B}\right)\sin(\alpha t)
\label{eq:EMfield-position-y}
\end{equation}
where $(x_0,y_0)$ and $(u_{x0},u_{y0})$ are the initial position and velocity. 
We have defined $\alpha=\frac{qB}{m}$, known as the 
cyclotron frequency. We now find out the focusing time $t_f$ from the solution $(x(t),y(t))$ following the same prescription described in the previous two examples. The focusing time $t_f$ is found, as before, by solving the equation
\begin{equation}
\left| \begin{array}{cccc}
1+\frac{1}{\alpha}\left (\frac{\theta_0}{2}+\sigma_{+0}\right ) \sin(\alpha t_f)  & & &  \frac{1}{\alpha}\left (\frac{\theta_0}{2}-\sigma_{+0}\right )(1-\cos(\alpha t_f)) \\ +\frac{1}{\alpha}\left (\sigma_{\times 0}-\omega_0 \right )(1-\cos(\alpha t_f))  & & &  +\frac{1}{\alpha}\left (\sigma_{\times 0}+\omega_0 \right ) \sin(\alpha t_f)\\  & & &   \\
-\frac{1}{\alpha}\left (\frac{\theta_0}{2}+\sigma_{+0}\right )(1-\cos(\alpha t_f)) & & & 1+\frac{1}{\alpha}\left (\frac{\theta_0}{2}-\sigma_{+0}\right ) \sin(\alpha t_f) \\ +\frac{1}{\alpha}\left (\sigma_{\times 0}-\omega_0 \right ) \sin(\alpha t_f) & & & -\frac{1}{\alpha}\left (\sigma_{\times 0}+\omega_0 \right )(1-\cos(\alpha t_f)) \end{array} \right| =0 
\end{equation}
The solution is
\begin{equation}
t_f=\frac{2}{\alpha}\tan^{-1}\left(-\frac{\alpha}{\theta_0 \pm 2\sqrt{I_0}}\right)
\label{eq:EMfield_focusing_time}
\end{equation}
It is clear one must have $I_0\geq0$ for $t_f$ to be real. For $I_0<0$, a real $t_f$ does not exist. We will now obtain 
the same results from the solutions of the ESR equation.

\noindent In this case, the ESR evolution equations take the following form
\begin{equation}
\frac{d\theta}{dt}+\frac{1}{2}\theta^2+2\left(\sigma_{+}^2+\sigma_{\times}^2-\bar{\omega}^2\right)+\frac{\alpha^2}{2}=0
\label{eq:EMfield-ESR1}
\end{equation}
\begin{equation}
\frac{d\sigma_{+}}{dt}+\theta\sigma_{+}-\alpha\sigma_{\times}=0
\label{eq:EMfield-ESR2}
\end{equation}
\begin{equation}
\frac{d\sigma_{\times}}{dt}+\theta\sigma_{\times}+\alpha\sigma_{+}=0
\label{eq:EMfield-ESR3}
\end{equation}
\begin{equation}
\frac{d\bar{\omega}}{dt}+\theta\bar{\omega}=0
\label{eq:EMfield-ESR4}
\end{equation}
where we have defined $\bar{\omega}=\omega-\frac{\alpha}{2}$. 
Defining $I=\sigma_{+}^2+\sigma_{\times}^2-\bar{\omega}^2$, one obtains 
Eqns. (\ref{eq:projectile-I1}) and (\ref{eq:projectile-I2}). Therefore, the 
scheme for solving these evolution equations is the same as before, though the structure of the equations for $\sigma_{+}$ and $\sigma_{\times}$ are 
different from that of the two dimensional oscillator. Here, 
the equation for $\sigma_{+}$ contains $\sigma_{\times}$ and that for 
$\sigma_{\times}$ contains $\sigma_{+}$. The solutions for the three cases 
are given by
\newline
1. $I>0$
\begin{equation}
\theta(t)=\frac{\alpha\sqrt{1+C^{2}}\cos(\alpha t+D)}{C+\sqrt{1+C^{2}}\sin(\alpha t+D)}
\label{eq:EMfield-I1-ESR1}
\end{equation}
\begin{equation}
\sigma_{+}(t)=\frac{E\sin(\alpha t)+F\cos(\alpha t)}{C+\sqrt{1+C^{2}}\sin(\alpha t+D)}
\label{eq:EMfield-I1-ESR2}
\end{equation}
\begin{equation}
\sigma_{\times}(t)=\frac{E\cos(\alpha t)-F\sin(\alpha t)}{C+\sqrt{1+C^{2}}\sin(\alpha t+D)}
\label{eq:EMfield-I1-ESR3}
\end{equation}
\begin{equation}
\omega(t)=\frac{\alpha}{2}+\frac{G}{C+\sqrt{1+C^{2}}\sin(\alpha t+D)}
\label{eq:EMfield-I1-ESR4}
\end{equation}
where
\begin{equation*}
C=\frac{\sqrt{I_0}}{\alpha}\left[-1+\frac{\theta_0^2}{4I_0}+\frac{\alpha^2}{4 I_0}\right]
\end{equation*}
\begin{equation*}
D=\tan^{-1}\left[\frac{2\sqrt{I_0}}{\theta_0}\left(\frac{\alpha}{2\sqrt{I_0}}-C\right)\right]
\end{equation*}
\begin{equation*}
\lbrace E,F,G \rbrace=\frac{\alpha}{2\sqrt{I_0}}\lbrace \sigma_{+0},\sigma_{\times 0},\omega_0-\frac{\alpha}{2} \rbrace
\end{equation*}
2. $I=0$
\begin{equation}
\theta(t)=\alpha \tan\left[\alpha(C-\frac{t}{2})\right]
\label{eq:EMfield-I2-ESR1}
\end{equation}
\begin{equation}
\sigma_{+}(t)=\sec^2\left[\alpha(C-\frac{t}{2})\right]\left[E\sin(\alpha t)+F\cos(\alpha t)\right]
\label{eq:EMfield-I2-ESR2}
\end{equation}
\begin{equation}
\sigma_{\times}(t)=\sec^2\left[\alpha(C-\frac{t}{2})\right]\left[E\cos(\alpha t)-F\sin(\alpha t)\right]
\label{eq:EMfield-I2-ESR3}
\end{equation}
\begin{equation}
\omega(t)=\frac{\alpha}{2}+G\sec^2\left[\alpha(C-\frac{t}{2})\right]
\label{eq:EMfield-I2-ESR4}
\end{equation}
where
\begin{equation*}
C=\frac{1}{\alpha}\tan^{-1}\left(\frac{\theta_0}{\alpha}\right)
\end{equation*}
\begin{equation*}
\lbrace E,F,G \rbrace=\frac{\lbrace \sigma_{+0},\sigma_{\times 0},\omega_0-\frac{\alpha}{2} \rbrace}{1+\left(\frac{\theta_0}{\alpha}\right)^2}
\end{equation*}
3. $I<0$
\begin{equation}
\theta(t)=\frac{\alpha C\cos(\alpha t+D)}{\sqrt{1+C^2}+C\sin(\alpha t+D)}
\label{eq:EMfield-I3-ESR1}
\end{equation}
\begin{equation}
\sigma_{+}(t)=\frac{E\sin(\alpha t)+F\cos(\alpha t)}{\sqrt{1+C^{2}}+C\sin(\alpha t+D)}
\label{eq:EMfield-I3-ESR2}
\end{equation}
\begin{equation}
\sigma_{\times}(t)=\frac{E\cos(\alpha t)-F\sin(\alpha t)}{\sqrt{1+C^{2}}+C\sin(\alpha t+D)}
\label{eq:EMfield-I3-ESR3}
\end{equation}
\begin{equation}
\omega(t)=\frac{\alpha}{2}+\frac{G}{\sqrt{1+C^{2}}+C\sin(\alpha t+D)}
\label{eq:EMfield-I3-ESR4}
\end{equation}
where
\begin{equation*}
C=\sqrt{-\frac{I_0}{\alpha^2}\left(1-\frac{\theta_0^2}{4I_0}-\frac{\alpha^2}{4 I_0}\right)^2-1}
\end{equation*}
\begin{equation*}
D=\tan^{-1}\left[\frac{2\sqrt{-I_0}}{\theta_0}\left(\frac{\alpha}{2\sqrt{-I_0}}-\sqrt{1+C^2}\right)\right]
\end{equation*}
\begin{equation*}
\lbrace E,F,G \rbrace=\frac{\alpha}{2\sqrt{-I_0}}\lbrace \sigma_{+0},\sigma_{\times 0},\omega_0-\frac{\alpha}{2} \rbrace
\end{equation*}

\noindent From the solutions, it is apparent that when $I\geq 0$, the trajectories 
meet in finite time; for $I<0$, they never meet. From (\ref{eq:EMfield-I1-ESR1}), one can show that focusing takes place, i.e.,
$\theta \to -\infty$ when $C+\sqrt{1+C^{2}}\sin(\alpha t_f+D)=0$. Thus it follows that the focusing time $t_f$ is the same as that given in (\ref{eq:EMfield_focusing_time}). Further, it may be 
noted that, nowhere in the solutions for the ESR variables, the electric field appears whereas the 
magnetic field appears through $\alpha$. Therefore, the evolution of the ESR 
variables with and without electric field are the same; the electric field only affects the equations of motion. Therefore, we take $E=0$ for simplicity. 
Unlike in the previous examples in two dimensions, 
in this case, the rotation is not identically zero even though the initial rotation is set to
zero. This happens because of the presence of the magnetic field through the parameter $\alpha$.

\begin{figure}[ht]
\centering
\subfigure[$\theta_{0}=2.0$, $\sigma_{+0}=0.0$, $\sigma_{\times0}=0.0$, $\omega_{0}=0.5$]{\includegraphics[scale=0.96]{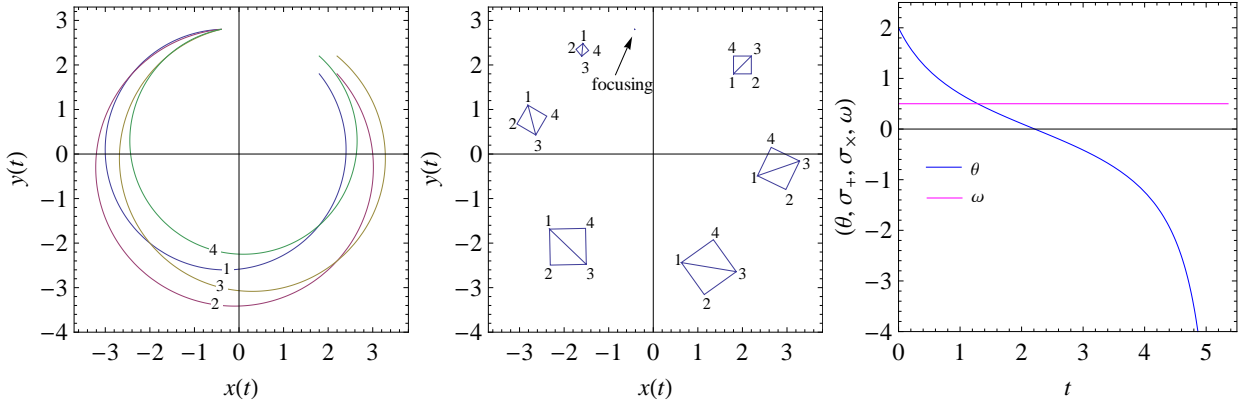}\label{fig:EMfield1a}}
\subfigure[$\theta_{0}=2.0$, $\sigma_{+0}=0.3$, $\sigma_{\times0}=0.4$, $\omega_{0}=0.0$]{\includegraphics[scale=0.96]{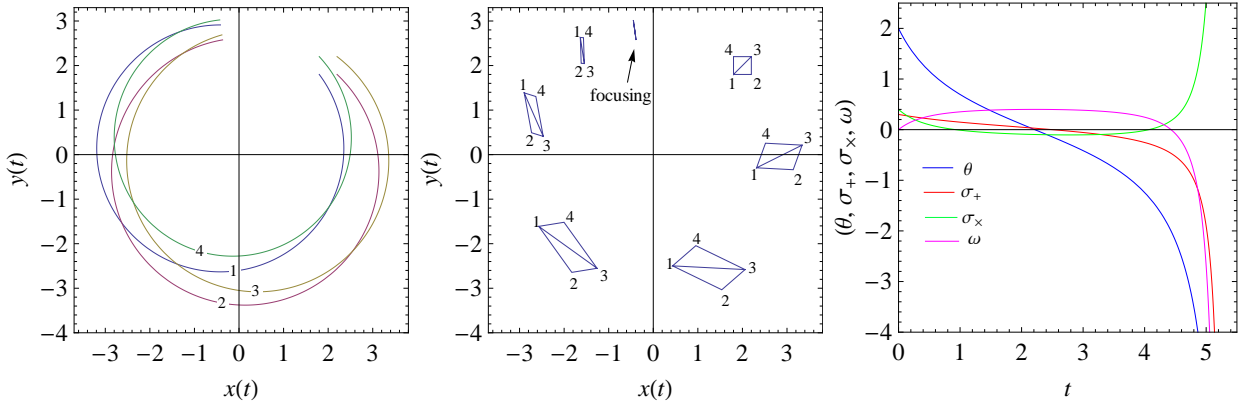}}
\caption{Plots for charged particle in magnetic field for $I=0$. The initial position and velocity of the central trajectory is $(2.0,2.0)$ and $(2.0,-2.0)$, respectively. Here, $\alpha=1$ and $B=1$.}
\label{fig:EMfield1}
\end{figure}

\begin{figure}[ht]
\centering
\subfigure[$\theta_{0}=2.0$, $\sigma_{+0}=0.3$, $\sigma_{\times0}=0.5$, $\omega_{0}=0.3$]
{\includegraphics[scale=0.96]{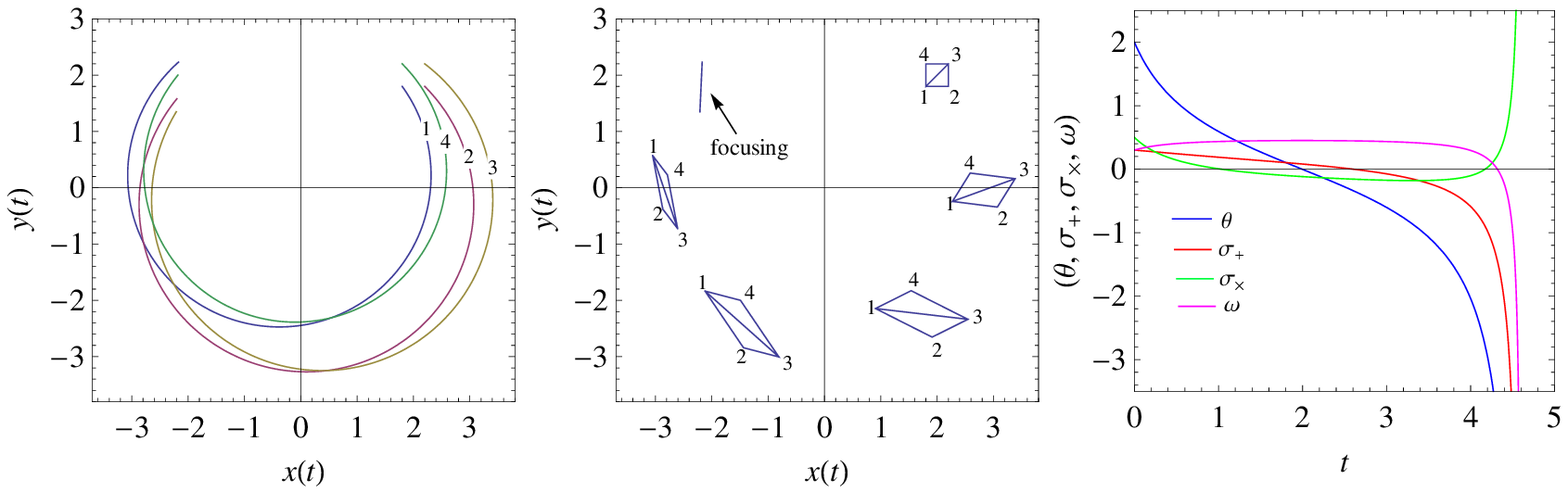}}
\caption{Plots for charged particle in magnetic field for $I>0$. The initial position and velocity of the central trajectory is $(2.0,2.0)$ and $(2.0,-2.0)$, respectively. Here, $\alpha=1$ and $B=1$.}
\label{fig:EMfield2}
\end{figure}

\begin{figure}[ht]
\centering
\subfigure[$\theta_{0}=2.0$, $\sigma_{+0}=0.0$, $\sigma_{\times0}=0.0$, $\omega_{0}=0.0$]{\includegraphics[scale=0.96]{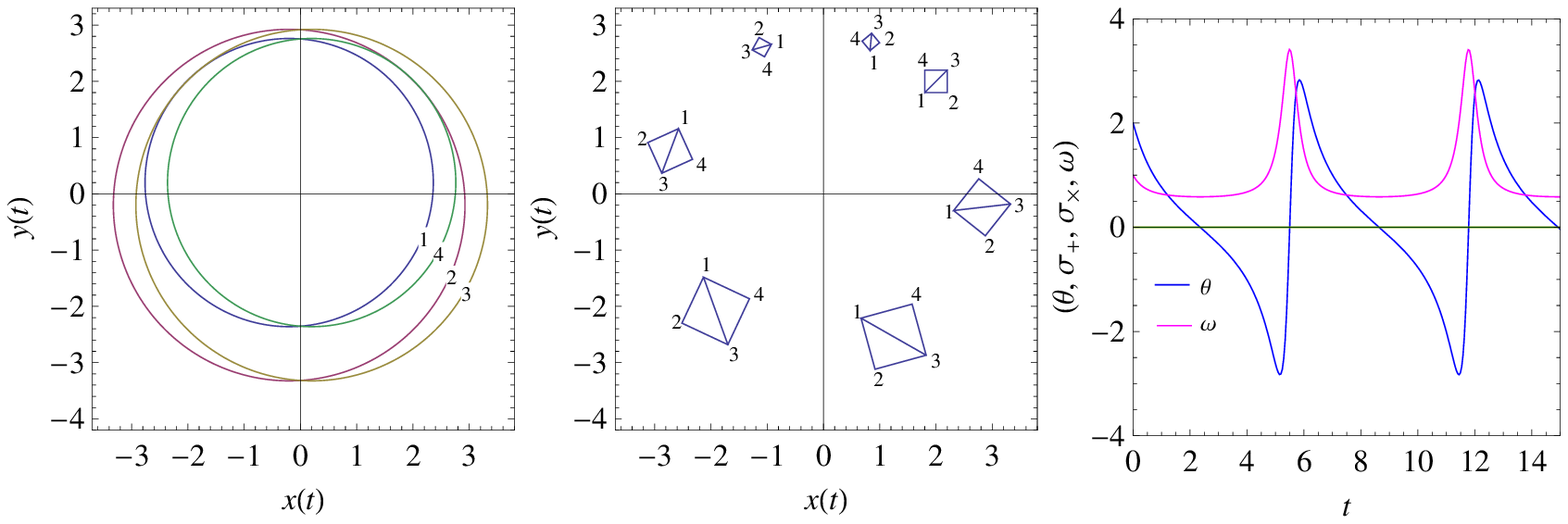}}
\subfigure[$\theta_{0}=2.0$, $\sigma_{+0}=0.5$, $\sigma_{\times0}=0.3$, $\omega_{0}=1.3$]{\includegraphics[scale=0.96]{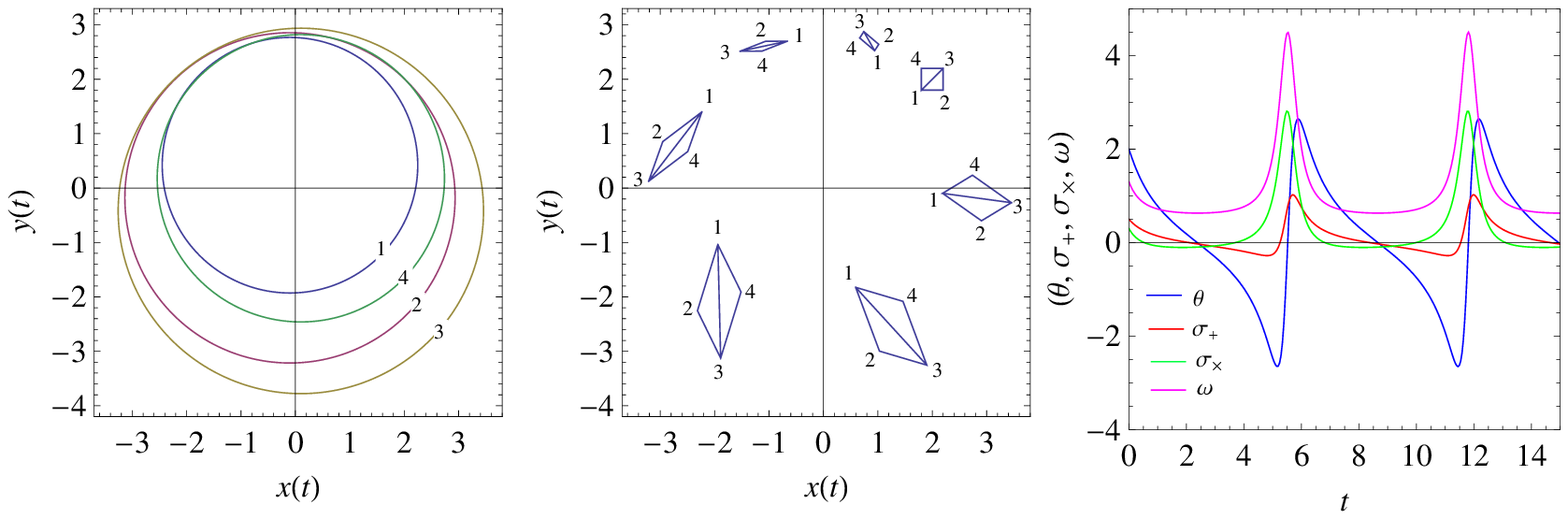}}
\caption{Plots for charged particle in magnetic field for $I<0$. The initial position and velocity of the central trajectory is $(2.0,2.0)$ and $(2.0,-2.0)$, respectively. Here, $\alpha=1$ and $B=1$.}
\label{fig:EMfield3}
\end{figure}

\noindent The evolution of a congruence of four particles starting from the four corners 
of a square, for different initial expansion, rotation and shear are shown in Fig.~\ref{fig:EMfield1}-\ref{fig:EMfield3}.
Once again the results shown in the figures demonstrate the role of
expansion, shear and rotation through the evolution of the shape of
the area enclosing the four trajectories. In this case, the nature of the evolution of the square is the same as 
that in the two dimensional oscillator case. 

\section{Connections}
\label{sec:connection}
\noindent As mentioned in Section \ref{approach}, the equations for the expansion, shear
and rotation which we have been dealing with are 
similar to the well-known Raychaudhuri equations
for geodesic congruences in a Riemannian or semi-Riemannian spacetime.
If $u^i$ is the velocity vector associated with the geodesic congruence, 
we can write down the following equation
\begin{equation*}
u^k\nabla_k(\nabla_j u^i)=u^k\nabla_j(\nabla_k u^i)+R^{li}_{\;\;jk}u_l u^k
\end{equation*}
\begin{equation}
\Rightarrow\hspace{0.2 cm}  u^k\nabla_k(\nabla_j u^i)=\nabla_j(u^k\nabla_k u^i)-(\nabla_j u^k)(\nabla_k u^i)+R^{li}_{\;\;jk}u_l u^k
\label{eq:ESRcurved}
\end{equation}
where $R^{li}_{\;\;jk}$ is the Riemann curvature tensor. On a Riemannian or
semi-Riemannian spacetime, the term $u^k\nabla_k u^i=0$ gives the geodesic 
equation for affinely parameterized geodesics. However,  for 
non-affinely parameterized geodesics or, say in Newtonian gravity, this term 
gives the force, i.e., acceleration of the particle, and hence 
is not equal to zero. Expressing the gradient of the velocity field, which
is a second rank tensor, as a sum of its trace, symmetric traceless and
antisymmetric parts (which essentially correspond to the expansion, shear
and rotation mentioned before) we can obtain individual equations for
these kinematic quantities. The resulting equations are known as the 
Raychaudhuri equations. More details on these equations are available
in \cite{hawk,wald,kar-sengupta}. We shall not discuss them any further here.
\section{What has been and can be done}
\label{sec:conclusion}
\noindent We have discussed the kinematics of a family of configuration space flow-lines of mechanical systems in one and two dimensions.
The kinematics has been quantified through the expansion, shear and rotation of a set of non-interacting particles set upon
trajectories started from perturbed initial conditions and visualized in the configuration space of the system.
This sheds some light on the general behaviour of mechanical systems in terms of the configuration space geometry. 

\noindent In more quantitative terms, the formalism presented here can
distinctly answer the following three questions for any given system.

$\bullet$ Is there a finite non-zero time at which a given set of
trajectories (for a particular physical system) with specified initial conditions, will meet? Is it possible to derive the conditions under which they
will never meet?

$\bullet$ What is the role of initial conditions on the behaviour of a given family of trajectories? 

$\bullet$ How does one develop a frame-by-frame (in time) picture of
the evolution of the family of trajectories? What are specific quantifiers of this evolution in time?

\noindent It will surely be worthwhile to study various mechanical systems in
diverse dimensions (especially in three dimensions) in order to
illustrate the formalism further and also to gain a better understanding.

\noindent In General Relativity, the expansion $\theta$ is known to
diverge to minus infinity (focusing) at a curvature singularity. However,
focusing can also occur in a benign way with the focal point not
being a curvature singularity but just a singularity of the congreunce. 
It is known that such congruence singularities do occur in optics (known
as caustics). For the classical mechanical
systems discussed here, focusing implies the meeting of trajectories.
It may be useful to know whether the focusing
of trajectories in mechanical systems can, in some situations, 
correspond to singularities in the geometry of configuration space 
or, alternatively, have an interpretation similar to the caustics in optics.

\noindent The approach advocated in this work is on the threshold of an immediate extension for investigating 
general dynamical systems which are extensively used as phenomenological 
models in a variety of areas in science and engineering. 
This sector is complex, largely unstructured, 
but affords exciting possibilities. 
The appearance of different flow structures and their transition in turbulent fluid flows
may be analysed using the proposed formalism. 
Focusing of a family of trajectories have important manifestations in, 
for example, shock formation
in fluid and solid mechanics, gravitational collapse in astrophysics and 
cosmology etc.. 
In the theory of dynamical systems, the issue of (non-metric) curvature of the underlying space and its meaning may be an 
interesting possibility to look into.

\noindent Looking ahead, the deeper question of integrability of dynamical systems may be associated with the evolution kinematics of the family of trajectories in the phase space \cite{arnold,mccauley}. It is known that if all flow-lines can be extended indefinitely (globally parallel flow), it provides a natural, globally Cartesian coordinate system implying integrability in the sense of Lie. On the other hand, occurrence of a singularity in the evolution of the family of trajectories implies an underlying effectively curved manifold which, in general, prevents integrability. The approach presented here, appropriately extended, may be able to provide some clue in this direction in future.

\section*{Acknowledgement}
\noindent Rajibul Shaikh acknowledges the Council of Scientific and Industrial Research (CSIR), India for providing support through a fellowship.


\begin{thebibliography}{99}
\bibitem{akr} A. Raychaudhuri, Phys. Rev. {\bf 98}, 1123 (1955).
\bibitem{hawk} S. W. Hawking and G. F. R. Ellis, {\em The large scale
structure of spacetime}, (Cambridge University Press, Cambridge, England, 1975).
\bibitem{wald} R. M. Wald, {\em General Relativity}, (University of Chicago Press, Chicago, USA, 1984).
\bibitem{kar-sengupta} S. Kar, S. Sengupta, Pramana {\bf 69}, pp. 49-76 (2007).
\bibitem{ADG1} A. Dasgupta, H. Nandan and S. Kar, Annals Phys. {\bf 323}, 1621 (2008).
\bibitem{ADG2} A. Dasgupta, H. Nandan and S. Kar, Int. J. Geom. Meth. Mod. Phys. {\bf 6}, 645 (2009).
\bibitem{ADG3} A. Dasgupta, H. Nandan and S. Kar, Phys. Rev. {\bf D79}, 124004 (2009).
\bibitem{SG} S. Ghosh, A. Dasgupta, S. Kar, Phys. Rev. {\bf D83}, 084001 (2011).
\bibitem{ADG4} A. Dasgupta, H. Nandan and S. Kar, Phys. Rev. {\bf D85}, 104037 (2012).
\bibitem{strogatz} S. H. Strogatz, {\em Nonlinear dynamics and Chaos : with applications to physics, biology, chemistry, and engineering}, (Perseus Books Publishing, Massachusetts, USA, 1994).
\bibitem{poisson} E. Poisson, {\em A relativist’s toolkit: the mathematics of black hole mechanics}, (Cambridge University Press, Cambridge, UK, 2004).
\bibitem{arnold} V.I. Arnold, {\em Mathematical Methods of Classical Mechanics}, (Springer-Verlag, 1989)
\bibitem{mccauley} J. McCauley, Chaos, Solitons \& Fractals, {\bf 5}(8), pp. 1493-1500 (1995).
\end{thebibliography}
\end{document}